\pdfoutput=1
\documentclass[11pt,twoside,a4paper,cmspaper,final,collab]{cms-tdr}
\def\svnVersion{1fee94c}\def\svnDate{2025/12/13}\def\cmsCernNoTag{CERN-EP-2025-041}\def\cmsCernDate{\today}\def\cmsMessage{Published in Physical Review D as \href{http://dx.doi.org/10.1103/rqvb-gfb7}{\doi{10.1103/rqvb-gfb7}.}}
\begin{document}\cmsNoteHeader{SMP-23-008}

\newcommand{\EPOS}{\textsc{epos-lhc}\xspace}
\newcommand{\pp}{\ensuremath{\Pp\Pp}\xspace}
\newcommand{\multifold}{\textsc{MultiFold}\xspace}
\newcommand{\cmsTable}[1]{\resizebox{\textwidth}{!}{#1}}
\newlength\cmsTabSkip\setlength\cmsTabSkip{1ex}
\cmsNoteHeader{SMP-23-008}
\title{Measurement of event shapes in minimum-bias events from proton-proton collisions at \texorpdfstring{$\sqrt{s} = 13\TeV$}{sqrt(s) = 13 TeV}}

\date{\today}
\abstract{A measurement of event-shape variables is presented, using a data sample produced in a special run with approximately one inelastic proton-proton collision per bunch crossing. The data were collected with the CMS detector at a center-of-mass energy of $13\TeV$, corresponding to an integrated luminosity of $64\mubinv$. A number of observables related to the overall distribution of charged particles in the collisions are corrected for detector effects and compared with simulations. Inclusive event-shape distributions, as well as differential distributions of event shapes as functions of charged-particle multiplicity, are studied. None of the models investigated are able to satisfactorily describe the data. Moreover, there are significant features common amongst all generator setups studied, particularly showing data being more isotropic than any of the simulations. Multidimensional unfolded distributions are provided, along with their correlations.}

\hypersetup{
pdfauthor={CMS Collaboration},
pdftitle={Measurement of event shapes in minimum-bias events from proton-proton collisions at \texorpdfstring{$\sqrt{s} = 13\TeV$}{sqrt(s) = 13 TeV}},
pdfsubject={CMS},
pdfkeywords={CMS, minimum bias, event shapes, instantons}}
\maketitle
\section{Introduction\label{sec:introduction}}

Recent measurements of observables in proton-proton (\pp) collisions related to the overall structure of the event, such as the correlations of outgoing particles~\cite{ALICE2021ridge,Khachatryan_2010,Khachatryan_2017,Khachatryan_2017b} and the multiplicity of strange hadrons~\cite{Strange2017,alicecollaboration2023studyingS}, highlight the difficulties in modeling these collisions. Models that attempt to describe these effects, for instance those developed in the context of heavy ion collisions, include nonperturbative effects, such as collective flow~\cite{Pierog_2015,Bartalini:2018qje,werner2023corecorona,werner2023parallel,werner2023deep,werner2023perturbative} or rope hadronization~\cite{bierlich2015effects,bierlich2016hadronisation,bierlich2016shoving}.  This motivates the continued measurement program of minimum-bias \pp collisions and the study of observables sensitive to nonperturbative effects. Furthermore, the dominance of soft quantum chromodynamics (QCD) effects, even in high-precision electroweak observables, such as the anomalous magnetic moment of the muon~\cite{Muong-2:2023cdq,Borsanyi_2021}, highlights the need for additional experimental input to nonperturbative QCD.

One nonperturbative phenomenon of particular interest is the QCD instanton~\cite{POLYAKOV197582,BELAVIN197585,PhysRevD.14.3432,PhysRevD.18.2199.3,PhysRevLett.37.172,CALLAN1976334,WITTEN1979269,Coleman1979,shuryak2020lectures}. Recent calculations~\cite{Khoze_2020} suggest that instanton processes at the CERN LHC may have a large cross section of $\mathcal{O}$(mb). Several strategies for detecting instanton events have been suggested~\cite{Khoze_2021,Khoze_2022,Tasevsky_2023,Amoroso_2021,shuryak2021observe}, and one of the common characteristics used in many of those searches is the relatively isotropic distribution of particles produced by the instanton. 

While we do not provide direct limits or instanton interpretation, these effects and other similar modifications of event shapes motivate our measurement as a means to test the limits of the existing \pp collision modeling. In particular, we study minimum-bias collisions that define a set of nearly inclusive collision events, with minimal selection criteria applied to ensure the collisions and outgoing particles are well measured by the detector.

This paper presents measurements of observables related to the overall kinematics of events, often referred to as event shapes. The measurements use minimum-bias \pp collisions at a center-of-mass energy of 13\TeV collected by the CMS detector at the CERN LHC, corresponding to an integrated luminosity of $64\mubinv$~\cite{LUM-17-003,LUM-18-002}. We study eight observables related to the global event shapes using tracks reconstructed in the detector (detector level). Then we unfold the same observables to the level of stable charged particles (particle level). Similar observables have been measured in $\Pep\Pem$ collisions at the CERN LEP collider~\cite{ALEPH:2003obs} and were compared with simulations in previous CMS event generator tuning studies~\cite{CMS:2019csb}. The measurements in this paper include effects of the hadronic initial state not present in those measurements. The charged-particle multiplicity has previously been studied at 13\TeV center-of-mass energy~\cite{CMS:2018nhd,ALICE:2022xip,ATLAS:2016zba}.  Some of the other observables have also been studied at the CERN LHC in the underlying event in top quark pair (\ttbar) production~\cite{Sirunyan_2019}, or using jets as inputs~\cite{ATLAS:2023mny}, and at lower center-of-mass energies~\cite{CMS:2011usu,ALICE:2012cor,ATLAS:2012uka}. The measurements presented here extend those analyses and elucidate correlations between the event shapes and the charged-particle multiplicity.
The observables are always computed using tracks (detector level) or charged particles (particle level). Sphericity, thrust, broadening and isotropy are computed in the center-of-mass frames of the tracks or charged particles for each event.
The full list of observables is as follows:

\begin{itemize}
    \item Particle multiplicity, $N$.
    \item Total invariant mass, $m_\text{ch} = \sqrt{(\sum_i p_i)^2}$, where $p_i$ is the four-momentum of particle $i$.
    \item Sphericity: a measure of how isotropically the momenta $p$ are distributed in an event~\cite{PARISI197865,PhysRevD.20.2759}. The tensor $S$ is first defined with components
    \begin{equation}
    \label{eq:sphericity}
    S^{\alpha\beta}=\frac{\sum_i p_i^\alpha p_i^\beta}{\sum_i\abs{\vec{p_i}}^2},
    \end{equation}
   in which $\alpha,\beta \in \{x,y,z\}$ refer to Cartesian coordinates, and $i$ is the index for the final-state charged particles that passed the selections based on the detector acceptance, described in Section~\ref{sec:selections}.  The sphericity is constructed from the two smallest eigenvalues $\lambda_2$ and $\lambda_3$: $\mathcal{S}=\frac{3}{2}(\lambda_2+\lambda_3)$.
    \item Thrust: a measure of how highly collimated the momenta in an event are along one particular axis, the thrust axis~\cite{PhysRevD.20.2759}. It is defined as 
    \begin{equation}
    \label{eq:thrust}
    \mathcal{T}=1-\max_{\vec{n}}\frac{\sum_i{\vec{p_i}}\cdot\vec{n}}{\sum_i\abs{\vec{p_i}}},
    \end{equation}
    where $\vec{n}$ is a unit vector, which is chosen to maximize the numerator~\cite{brandt1979axes}.
    \item Transverse thrust: the thrust in the transverse plane, calculated by setting $p_i^z$ to 0 in the thrust definition (Eq.~\ref{eq:thrust}).
    \item Broadening: a measure of the fraction of energy that is perpendicular to the thrust axis. The thrust axis defines the left $\mathcal{L}$ and right $\mathcal{R}$ hemispheres of the event. The left and right broadening are defined as 
    \begin{equation}
    \mathcal{B_L}=\sum_{i\in\mathcal{L}}\frac{\abs{\vec{p_i}\times\vec{n}}}{2\sum_j\abs{\vec{p_j}}},\ \mathcal{B_R}=\sum_{i\in\mathcal{R}}\frac{\abs{\vec{p_i}\times\vec{n}}}{2\sum_j\abs{\vec{p_j}}}.
    \end{equation}
    The total broadening is defined as $\mathcal{B}=\mathcal{B_L}+\mathcal{B_R}$.
    \item Transverse spherocity: the spherocity in the transverse plane, a measure of how spread out the momentum is across directions in the transverse plane,
    \begin{equation}
    \mathcal{S}_0 = \frac{\pi^2}{4} \min_{\vec{n} = (n_x, n_y, 0) } \left(\frac{\sum_i \abs{\vec{\pt}_i \times \vec{n}}}{\sum_i {\pt}_i}\right)^2.
    \end{equation}
    \item Isotropy: a measure of how isotropically energy is distributed in an event. It is defined as the energy mover's distance~\cite{komiske2019energy} from the spatial distribution of the tracks to a uniform radiation pattern~\cite{cesarotti2020robust}. The detailed definition of this variable is given in Ref.~\cite{cesarotti2021efficacy}, which also explores the efficacy of the isotropy to distinguish the isotropic events from the jetlike ones characterized by the majority of the transverse momentum being clustered along a few axes. For this measurement we use the spherical geometry, with the distance measure $d_{ij} = 2(1-\cos\theta_{ij})$, where $\theta_{ij}$ is the angle between the particles.  
\end{itemize}

We investigate a number of minimum-bias simulations from generators including \HERWIG (v7.2.2)~\cite{B_hr_2008,Bellm:2015jjp}, \EPOS~\cite{Pierog_2015}, and \PYTHIA (v8.230)~\cite{Sjostrand:2014zea}, with multiple different tunes of \PYTHIA, to evaluate the current quality of the modeling of event-shape distributions, and to provide a reference for the future search for currently unmodeled instanton contributions to these observables. 

A machine learning (ML)-based unfolding algorithm~\cite{andreassen2020omnifold,andreassen2021scaffolding} is used to correct for detector effects in the observed distributions and to estimate the distributions for stable charged particles. This ML-based approach is a multidimensional unfolding, which follows a similar approach to traditional binned iterative unfolding, but extended to an unbinned analysis.

For each event, the value of every unfolded observable is provided and a simultaneous unfolding of all observables is performed.  This allows the full joint information of all unfolded observables to be taken into account simultaneously in the unfolding and for further selections or binning choices to be made after the unfolding is performed.

The histograms of individual event-shape observables are provided in Section~\ref{sec:results}, as well as two-dimensional (2D) histograms of event-shape observables as functions of the number of charged particles and their invariant mass in Section~\ref{sec:results} and Appendix~\ref{sec:other-2d-distributions}. Correlation matrices for the systematic uncertainties of the histogrammed event-shape observables are provided in Appendix~\ref{sec:correlation}. Tabulated results are provided in the HEPData record~\cite{hepdata}, and the unbinned unfolding results are provided in the CERN Document Server~\cite{CDSdata}.

\section{The CMS detector and object reconstruction\label{sec:cms_and_objects}}

The central feature of the CMS apparatus is a superconducting solenoid of 6\unit{m} internal diameter, providing a magnetic field strength of 3.8\unit{T}. Within the solenoid volume are a silicon pixel and strip tracker, a lead tungstate crystal electromagnetic calorimeter (ECAL), and a brass and scintillator hadron calorimeter (HCAL), each composed of a barrel and two end-cap sections. Forward calorimeters extend the pseudorapidity coverage provided by the barrel and end-cap detectors. Muons are measured in gas-ionization detectors embedded in the steel flux-return yoke outside the solenoid. A more detailed description of the CMS detector, together with a definition of the coordinate system used and the relevant kinematic variables, can be found in Refs.~\cite{CMS:2008xjf,CMS:2023gfb}. 

The primary vertices (PVs) are estimates of the positions of the \pp collisions reconstructed from the detector objects. We use the standard CMS vertex reconstruction algorithm~\cite{cms2014description}, which clusters prompt tracks, associates them with PVs, and then fits the PV positions from the tracks. A particle-flow algorithm~\cite{CMS:2017yfk} aims to reconstruct and identify each individual particle in an event, with an optimized combination of all subdetector information. In this process, the identification of the particle type (photon, electron, muon, charged hadron, or neutral hadron) plays an important role in the determination of the particle direction and energy. Electrons (\eg, coming from photon conversions in the tracker material or from \PB meson semileptonic decays) are identified as a primary charged-particle track and potentially many ECAL energy clusters corresponding to this track followed to the ECAL and to possible bremsstrahlung photons emitted along the way through the tracker material. Muons (\eg, from \PB meson semileptonic decays) are identified as tracks in the central tracker consistent with either a track or several hits in the muon system, and associated with calorimeter deposits compatible with the muon hypothesis. Charged hadrons are identified as charged-particle tracks, neither identified as electrons nor as muons. 

The energy of electrons is determined from a combination of the track momentum at the main interaction vertex, the corresponding ECAL cluster energy, and the energy sum of all bremsstrahlung photons attached to the track. The energy of muons is obtained from the corresponding track momentum. The energy of charged hadrons is determined from a combination of the track momentum and the corresponding ECAL and HCAL energies, corrected for the response function of the calorimeters to hadronic showers. 

\section{Data and simulated samples\label{sec:samples}}

Data were collected with the CMS detector in 2018 under low-pileup conditions, corresponding to an average number of inelastic $\pp$ collisions per bunch crossing of approximately 1. The usage of the low-pileup data mitigates pileup effects in this precision measurement~\cite{CMS:2020ebo}. Despite the low luminosity of this dataset, the measurement is not statistically limited. Only a fraction of the total data collected by the CMS detector in 2018 is used, corresponding to 7 million events and an integrated luminosity of $64\mubinv$~\cite{LUM-17-003,LUM-18-002}. An unbiased trigger is used to randomly select collision events that are recorded independently of the event content~\cite{Khachatryan_2017Trigger}.

Several Monte Carlo (MC) simulations of minimum-bias \pp collisions are used to perform the unfolding and compare with the observed data. One nominal choice of minimum-bias event generation setup is used as the central prediction, while the others are used for estimating uncertainties as well as for comparison and interpretation. Some details of the nominal simulation sample and of those used for estimating systematic uncertainties are listed in Table~\ref{table:simulated_samples}.
In all cases, \GEANTfour~\cite{Agostinelli:2002hh} is used for the detector simulation. For most samples, approximately 2 million unweighted events are used.

The \PYTHIA CP1 tune~\cite{CMS:2019csb} is chosen as the nominal model for its relatively good description of the observed distributions of track multiplicity and track pseudorapidity to minimize uncertainties and bias in the unfolding due to mismodeling. The \PYTHIA A3 tune~\cite{ATL-PHYS-PUB-2016-017}, \EPOS~\cite{Pierog_2015}, and the \HERWIG CH3 tune~\cite{sirunyan2021development} were chosen as systematic variations that provide a similar level of agreement with the observations, while also covering different tuning and physics model choices. \PYTHIA and \HERWIG are both general purpose MC event generators commonly used to perform \pp collision analyses. \EPOS uses a core-corona event model for hadronic collisions originally developed in the context of heavy ion collisions but adapted for \pp collisions and includes some LHC data in the parameter tuning.

\begin{table*}[htbp!]
    \centering
    \topcaption{Samples of simulated minimum-bias events used for the unfolding. The value of the strong coupling constant, $\alpS$, corresponds to that used in the parton distribution function (PDF) set. The samples are simulated without additional pileup interactions.}
    \begin{scotch}{lll}
     & Generator & PDF ($\alpS(m_{\PZ})$) \\
    \hline
    Nominal sample       & \PYTHIA{8} (CP1)~\cite{CMS:2019csb} & NNPDF3.1 QCD LO  (0.130) \\[\cmsTabSkip]
    Systematic variations &  \EPOS~\cite{Pierog_2015} &   \\
                                        & \PYTHIA{8} (A3)~\cite{ATL-PHYS-PUB-2016-017} &  NNPDF2.3 QCD+QED LO  (0.130)  \\
                                        & \HERWIG 7 (CH3)~\cite{sirunyan2021development} & NNPDF3.1 QCD NNLO  (0.118) \\
    \end{scotch}
    \label{table:simulated_samples}
\end{table*}

Additional MC simulations are listed in Appendix~\ref{sec:additional_samples}, which are used for cross-checks and validation but are not included in the main analysis. Some of these samples were used as pseudodata samples to validate the performance of the unfolding against known particle-level distributions, while others were investigated to ensure that they did not give substantially different results to the samples used in the analysis. The main samples used in the analysis are all simulated without additional pileup interactions. A simulated \PYTHIA sample with the CP5 tune with pileup drawn from a Poisson distribution of mean 1 was used to check the impact of pileup on the observables--the effects were negligible. 

\section{Event and object selections\label{sec:selections}}
This analysis is performed using charged particles (primarily charged hadrons) without clustering into jets. All calculations are performed directly with tracks associated with particle-flow candidates and reconstructed vertices without distinguishing leptons and hadrons. 
    
The analysis uses a minimal set of selections at detector level, applied to both data and simulation, to select events with one inelastic proton-proton collision. A selection is made on the PV, such that it is within a broad region around the nominal interaction point: within ${\pm}24\unit{cm}$ along the beamline direction and within ${\pm}2\unit{cm}$ in the transverse plane. Furthermore, the vertex position must be well reconstructed from the tracks associated with it, as determined by the number and position of tracks associated with the vertex~\cite{cms2014description}.  Every event is required to have only one such PV and at least three tracks fulfilling the track selection criteria associated with the PV. All tracks are required to have transverse momentum $\pt > 0.5\GeV$, pseudorapidity $\abs{\eta}<2.4$, and loose selections are placed on the track impact parameters based on the vertex resolution: the track impact parameter in the transverse (longitudinal) direction must be at most 8 (13) times the vertex resolution in the transverse (longitudinal) direction.
    
A minimal particle-level event selection defines the fiducial phase space that is the target of the unfolding. These selections are intended to define a minimally biased set of inelastic collisions, while minimizing extrapolation between the detector-level selection and the particle-level phase space. Events are required to have at least two charged particles with $\pt > 0.5\GeV$ and $\abs{\eta} < 2.4$. The particles selected at the particle level are electrically charged particles and required to be stable (lifetime $c\tau > 10\unit{mm}$, corresponding to ``status 1" particles of the \textsc{HepMC} standard~\cite{Buckley_2021}).  Simulated events that pass the detector-level selection but do not meet the fiducial phase space selection criteria are included in the analysis and treated as background.
    
\section{Unfolding procedure\label{sec:unfolding}}

We use an ML-based unfolding procedure, \multifold, which performs an unbinned, multidimensional unfolding and uses iteration as its method of regularization~\cite{andreassen2020omnifold,andreassen2021scaffolding}.
    
The \multifold algorithm is similar to the Richardson-Lucy unfolding~\cite{Richardson:72,1974AJ79745L}, also known as iterative Bayesian unfolding or the d'Agostini method~\cite{DAGOSTINI1995487}, but uses ML to extend the methods to unbinned cases. The algorithm is based on the fact that, when performing a classification to identify an event as coming from one sample or another, an ideal classifier will learn the likelihood ratio, \ie the ratio of the probability of observing such an event given that it is from one class or the other. Therefore, a sufficiently well-designed and trained neural network classifier should approximate the likelihood ratio~\cite{Cranmer:2015bka}. The likelihood ratio estimated from the classifier is then used to reweight events from one sample to resemble the other sample in an unbinned way. 
    
The unfolding uses multiple reweighting steps for each iteration, first correcting the simulation toward the data at the detector level, then correcting the original simulation toward this corrected simulation at the particle level. Additional steps are also employed to account for acceptance and efficiency effects, as described in Ref.~\cite{andreassen2021scaffolding}. 
    
These steps are iterated a number of times, with the iteration acting as a form of regularization. Fewer iterations correspond to more regularization and tend to be more biased toward the input simulated distribution. The number of iterations used in the unfolding was chosen by checking goodness-of-fit measures of the unfolded result on a simulated pseudodata sample from an alternative simulation. The goodness of fit of the histograms reported in Section~\ref{sec:results} was calculated using a $\chi^2$ for each observable, and the number of iterations was chosen such that larger numbers of iterations did not provide any clear improvement in the goodness of fit. Both the bias of the unfolded estimates and the frequency with which the stated uncertainty intervals contain the true value of the parameter being estimated (coverage) converge quickly, and no significant changes were found to these properties after two iterations.
    
For every event, the neural networks are provided with the values of all observables being unfolded, such that all observables are simultaneously unfolded, and their correlations are taken into account. The values passed for every event are the number of objects, their invariant mass, sphericity, thrust, transverse thrust, broadening, transverse spherocity, and isotropy, as defined in Section~\ref{sec:introduction}. The objects used to calculate the observables are the tracks for detector-level reweighting, or charged particles for particle-level reweighting.
    
The classification networks consist of three fully connected layers of 100 nodes each. The cross-entropy, which measures the ability of the network to identify the sample from which an event originates, is used as the loss function. The training is performed for every step of the procedure and every iteration, and is done using 80\% of the sample for training and 20\% for validation. Early stopping is used, where the training is stopped after no improvement in the loss function evaluated on the validation sample is seen for 10 epochs (complete passes through the training sample). Then the network parameters are taken from the last epoch that showed improvement in the training. Different initial values of the network weights during training can lead to random fluctuations in the output. This effect was mitigated by an ensembling technique: four trainings and evaluations with independent initializations of the nerual network weights are performed, and the final event weight is the average of the ensemble. The unfolding is performed using two iterations.
    
As an input to the unfolding, the nominal MC simulated events have uniform weights that normalize them to the total inelastic cross section of 13\TeV \pp collisions reported by CMS \cite{sirunyan2018measurement}.  All MC samples are weighted such that the sum of weights for the events passing the particle-level selection is equal to that for the nominal MC sample. The weighting ensures a fixed fiducial cross section for the nominal MC and its variations at the start of the unfolding. 
    
In each reweighting step of the training, the program uses balanced weights between the set to be weighted and the target, so that the weighting function from the ML model is based on only the shape differences between the distributions of the two sets. The weights from the unfolding are multiplied by the previous ones in each iteration.
    
The final result of the unbinned unfolding is a set of reweighted MC events, whose weights provide an estimate of the probability density of the observed data at each point in the phase space. These results are binned after the unfolding to provide numerous histograms displaying the estimated differential cross sections as a function of the observables of interest.
    
A number of tests have been performed to validate the unfolding procedure. For these tests, one or more of the alternative simulated samples listed in Table~\ref{table:additional_samples} is treated as pseudodata to be unfolded. As well as visual checks of the unfolding closure, binned goodness-of-fit tests are performed to validate that the unfolded results provide a better description of the pseudodata than the original simulation. A bottom-line test~\cite{cousins2016unfolded} is performed to ensure that the unfolding does not artificially introduce discriminating power between models that was not already present at the detector level. Frequentist bias and coverage tests were also performed using parametric sampling~\cite{LEE2005143}. For most bins, the estimated bias is compatible with zero, and in all cases it is considerably smaller than the overall uncertainty. No cases of undercoverage were observed.
    
\section{Uncertainties\label{sec:uncertainties}}

The statistical uncertainty in the observed data is estimated by a bootstrapping method \cite{tibshirani1993introduction}. This method first assigns weights following a Poisson(1) distribution to the data events to generate alternative data samples with statistical fluctuations, and then performs pseudoexperiments of unfolding these alternative data samples. The standard deviations of the unfolded results become estimations of the statistical uncertainty.
    
The effects of systematic uncertainties are also estimated using pseudoexperiments, which start from templates defining the shifts in the unfolding model from individual systematic uncertainty sources. 
The templates define these shifts as weights on the events of the nominal MC sample.
The sources of the systematic uncertainty and the estimation methods are as follows:
\begin{itemize}
    \item MC modeling:
    modeling of the underlying physics can affect the result of the unfolded measurements in several distinct ways. We consider two distinct effects:
    \begin{itemize}
        \item The regularization, which is induced by the limited number of iterations in the unfolding algorithm and imperfections in the neural network classifier performance can lead to biased results. This bias depends on the choice of the nominal MC model. To estimate the effect of this systematic uncertainty, alternative templates are estimated by weighting the nominal MC to the alternative MC samples from other tunes or generators at the particle level. The resulting reweighted nominal sample has the same particle-level distributions as the alternative MC but maintains the same migration function as the nominal ones.
        \item Differences in the underlying physics modeling of quantities that are not unfolded, for example the track pseudorapidity distribution, can also affect the detector response and thus lead to a poor modeling of the migration from the particle-level distributions to the detector-level ones. The templates of this systematic deviation are derived in two steps. In the first step, the alternative MC samples are weighted to the nominal sample at the particle level, which results in reweighted alternative MC samples with the same particle-level distributions as the nominal MC and the same migration function as the alternative MC. In the second step, the nominal sample is weighted at both the particle level and the detector level to the reweighted alternative MC samples from the first step. The second step results in reweighted nominal samples with the same particle-level and detector-level distributions as the output from the first step. The function of the second step is to transfer the weights on the alternative MC samples to those on the nominal samples, keeping the distributions the same. The resulting reweighted nominal samples have the same particle-level distribution as the nominal MC and the same migration function as the alternative MC.
    \end{itemize}
    For each of these two effects, we consider multiple alternative MC models as sources of uncertainty. Specifically, we consider one alternative \PYTHIA tune as well as the \EPOS and \HERWIG setups, as described in Section~\ref{sec:samples}. Therefore, we have a total of six independent modeling uncertainties. Alternative choices of systematic samples coming from generator tune eigenvariations or other models with reasonable descriptions of the data were also evaluated and they resulted in similar sizes of the uncertainties.
    \item Track reconstruction efficiency: this uncertainty accounts for potential deviations of the track reconstruction efficiency in simulations from the actual efficiency in observations.  To estimate the effect of this uncertainty in the event-shape measurement, we randomly drop 2.1\% of the tracks with $\pt<20\GeV$ and 1\% of those with $\pt>20\GeV$ in the simulation. These numbers are based on the tracking efficiency studies in the $\mathrm{D}^*(2010)^\pm$ meson reconstruction measurement \cite{trackeff,cms2712742muon}.  The template of this systematic deviation is estimated by weighting the nominal MC samples to the samples with randomly dropped tracks at the particle and detector levels. 
    \item Statistical uncertainty of the simulated model:
        the statistical uncertainty of the MC sample leads to a systematic uncertainty in the unfolding results. This uncertainty is estimated by the same bootstrapping method that is used in estimating the uncertainty from the event counts of observed data. 
\end{itemize}
    
The effects of pileup were negligible.
    
For the MC regularization bias, migration function, and track reconstruction uncertainties, we derived a set of weights on the nominal MC simulation, such that the effect of the uncertainty can be described as a function of a continuous nuisance parameter. The weights are derived from the alternative samples using a ML-reweighting procedure similar to the reweighting procedure used in the unfolding. A neural network classifier is trained to distinguish between the nominal and alternative samples and then the classifier score is used to define the reweighting value. The nuisance parameters for each uncertainty are assumed to be Gaussian-constrained, and the weight of any event, $w_i$, as a function of a given nuisance parameter, $\theta$, is given by
\begin{equation}
    w_i(\theta) =  w_\text{nom}\, (w_i(1)/w_\text{nom})^{\theta},
\end{equation}
in which $w_\text{nom}$ is the weight of the nominal MC event, and $w_i(1)$ is the weight of the event in the corresponding systematic uncertainty template, which is defined to be the +1 standard deviation from the nominal.
    
The total uncertainty and its covariance matrices are estimated by 100 pseudoexperiments of unfolding with weighted nominal MC simulation and weighted data separately, which emulate the fluctuations of the templates. For the data and MC statistical uncertainty, the weights in the pseudoexperiments are taken from the Poisson(1) distribution and applied to the data and MC samples, respectively. For the MC bias, migration function, and track reconstruction efficiency uncertainties, a nuisance parameter value $\theta_j \sim N(0,1)$ is generated for each template $j$. The total weight of any given MC event for each pseudoexperiment is the product of the weights from all the sources of systematic uncertainty. 
    
The systematic uncertainties in the results are larger than the statistical uncertainties for the chosen bin widths in this paper. For one-dimensional (1D) distributions, the typical uncertainty is $\mathcal{O}(5\%)$, depending on the kinematic bin.  The uncertainty from tracking efficiency is typically the largest, of the order of a few percent. The uncertainty from the migration functions for different physics models combined across all models is of a similar size. The uncertainty from the estimated regularization bias between the nominal model and other models is typically $\mathcal{O}(1\%)$ or less for each model, and the combined uncertainty from regularization bias from all models  is typically a little smaller than the uncertainties from tracking efficiency or migration functions of the models. For the bin widths presented here, the statistical uncertainties in the 1D distributions are significantly less than 1\%.
    
\section{Results}\label{sec:results}

Projections of the full unfolded distribution into 1D and 2D histograms are provided and discussed below. Additional 2D histograms and the correlation matrices of the systematic uncertainties for the 1D projections are given in Appendices~\ref{sec:other-2d-distributions} and \ref{sec:correlation}. The uncertainty decomposition of the unfolding is given in Appendix~\ref{sec:uncertainty_decomposition}.
    
\begin{figure*}[!ht]
    \includegraphics[width=0.48\textwidth]{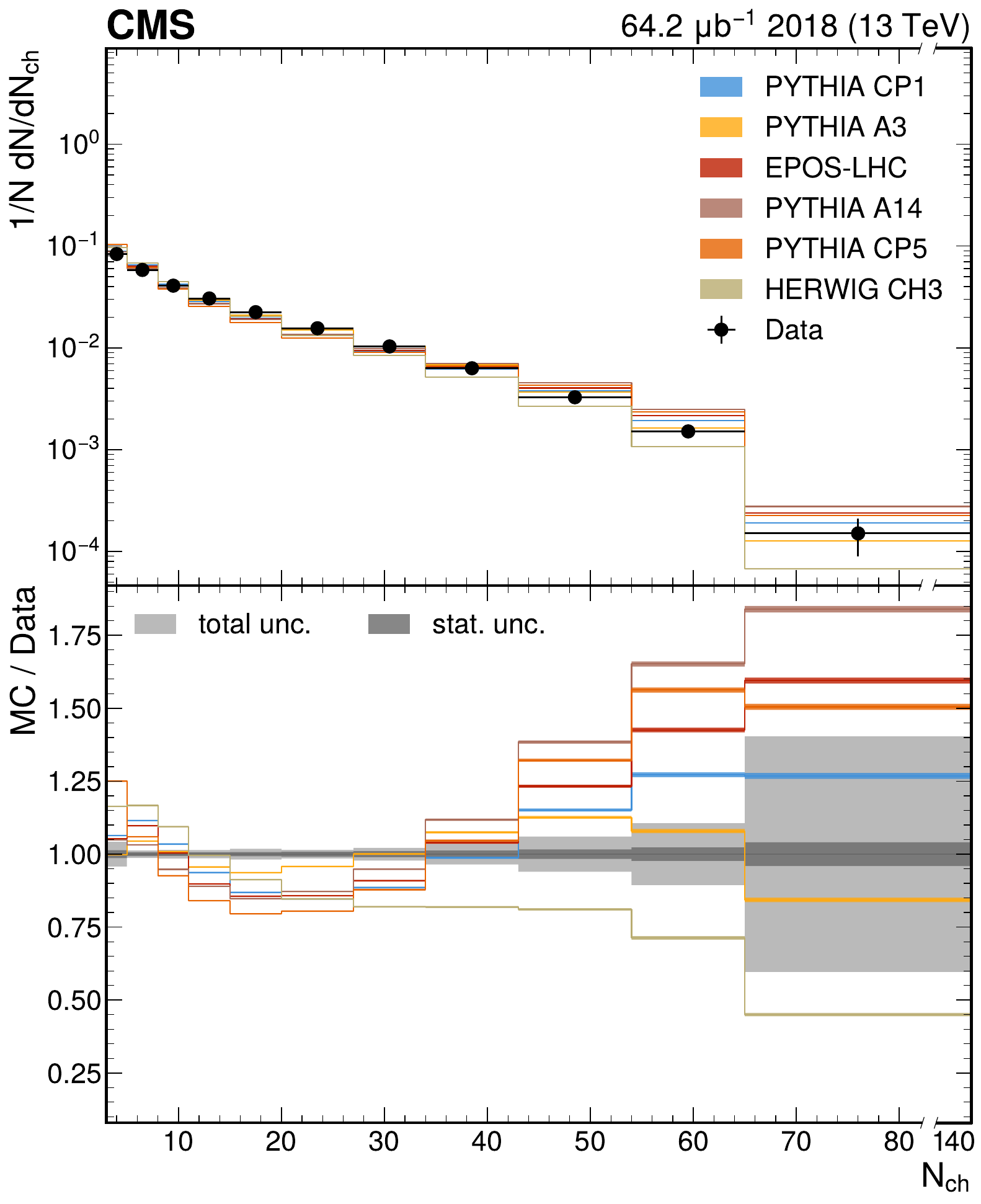}
    \includegraphics[width=0.48\textwidth]{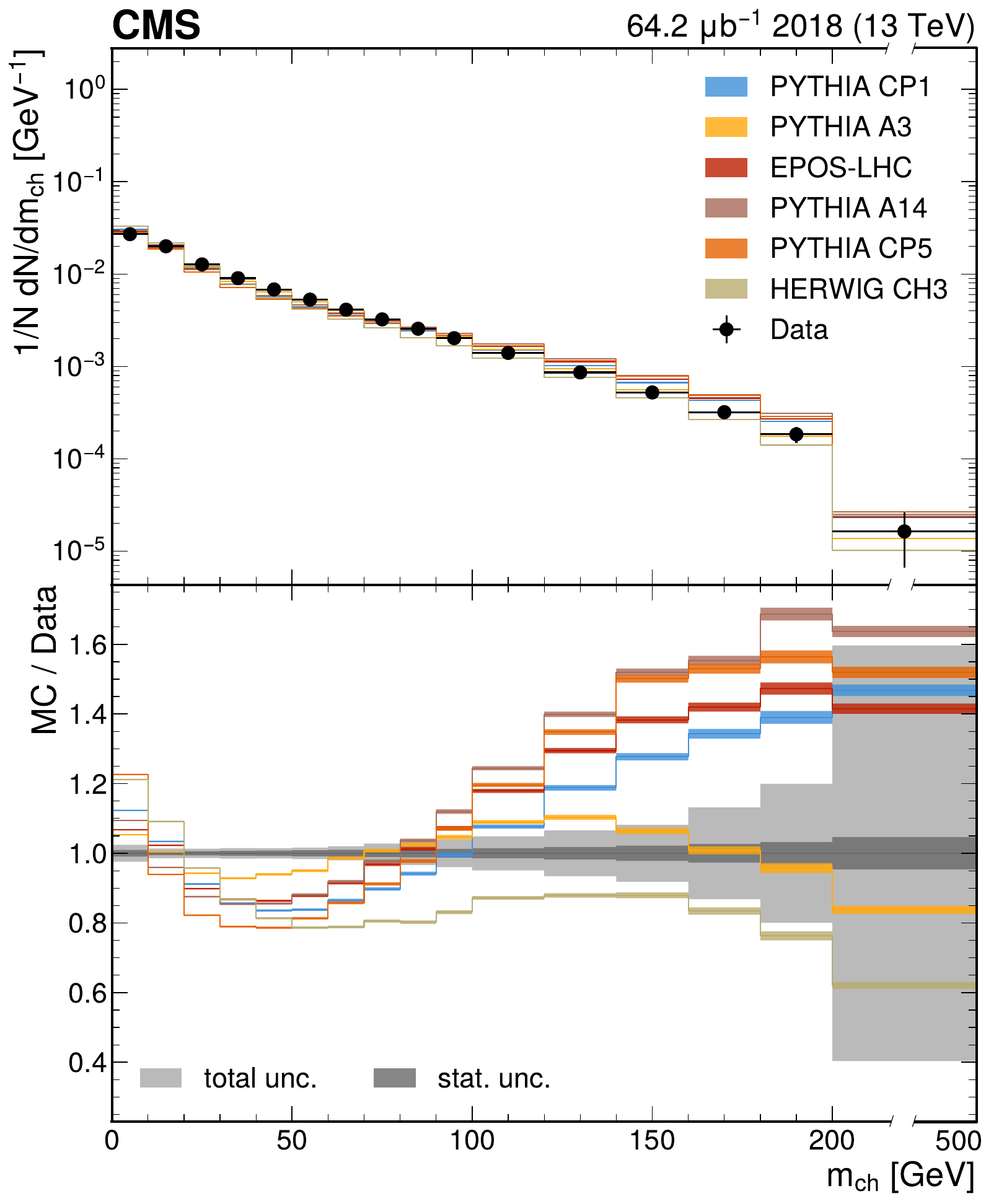}
    \caption{The unfolded results for (left) the particle multiplicity, and (right) invariant mass, compared with the nominal MC from the \PYTHIA CP1 tune and predictions from the \PYTHIA A14, CP5, A3 tunes, the \EPOS generator, and the \HERWIG CH3 tune. The statistical uncertainty on each of the predictions is shown by its shaded band.}
    \label{fig:unfold-gen-1d-1}
\end{figure*}

\begin{figure*}[!ht]
    \centering
    \includegraphics[width=0.48\textwidth]{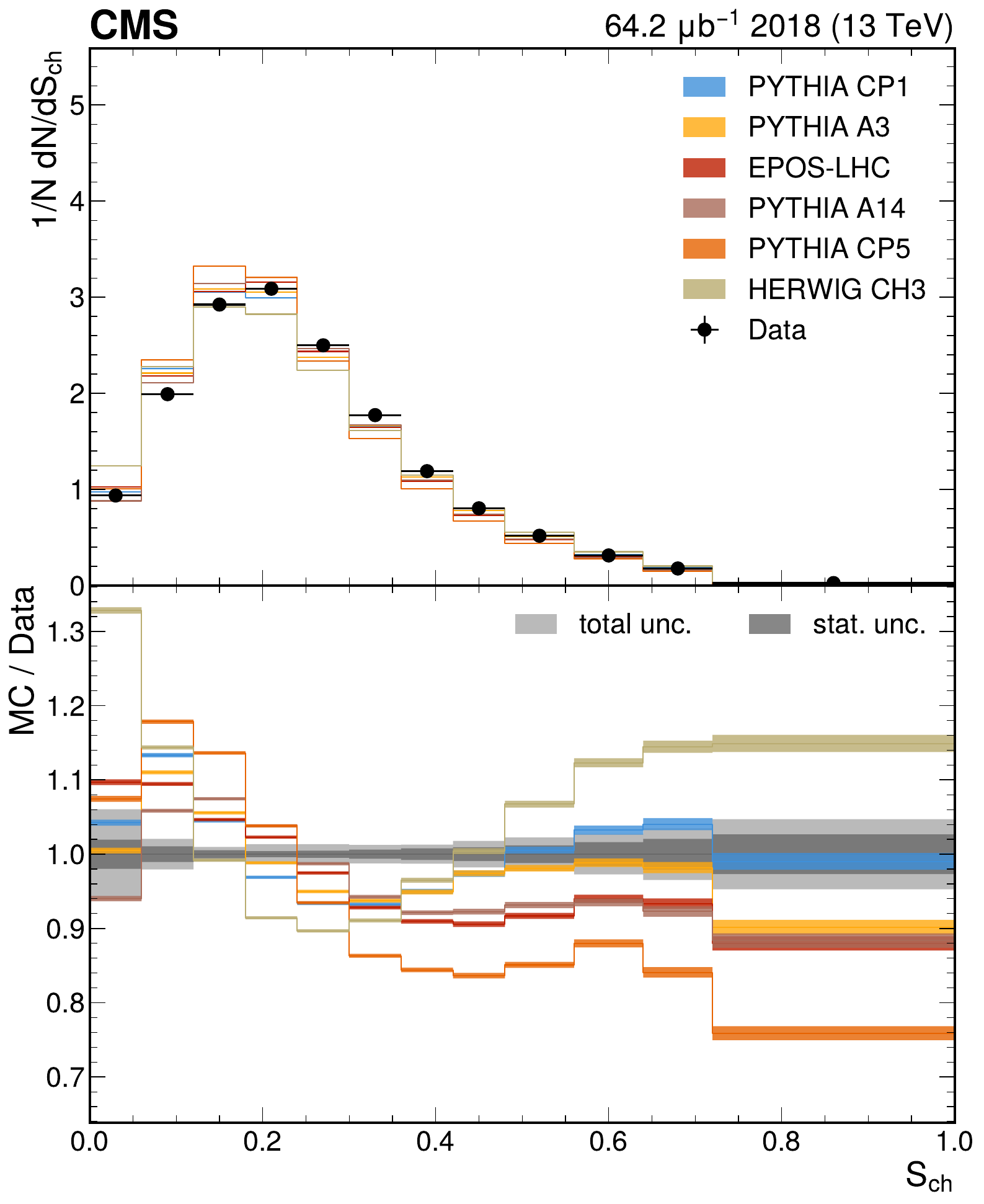}
    \includegraphics[width=0.48\textwidth]{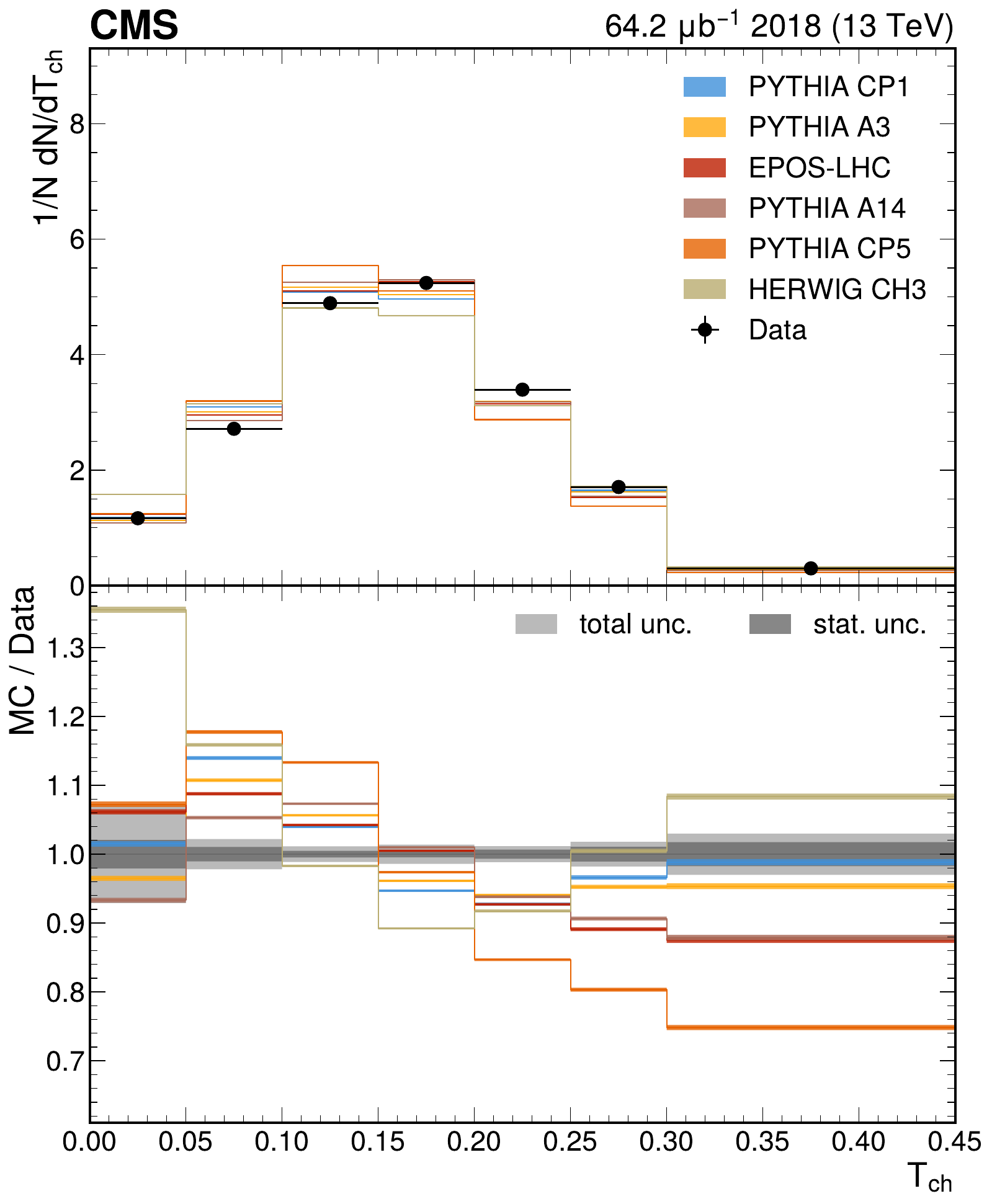}
    \includegraphics[width=0.48\textwidth]{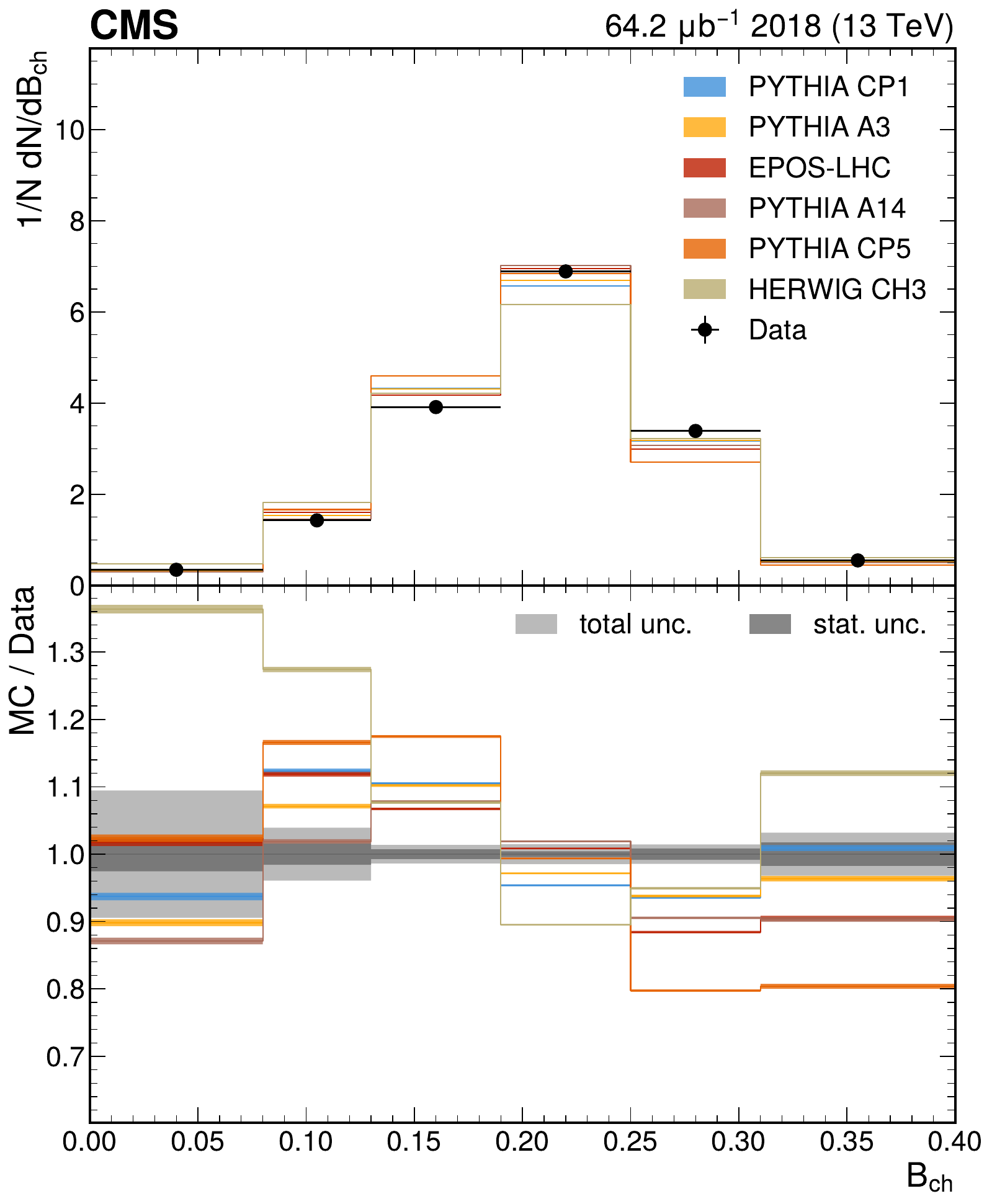}
    \includegraphics[width=0.48\textwidth]{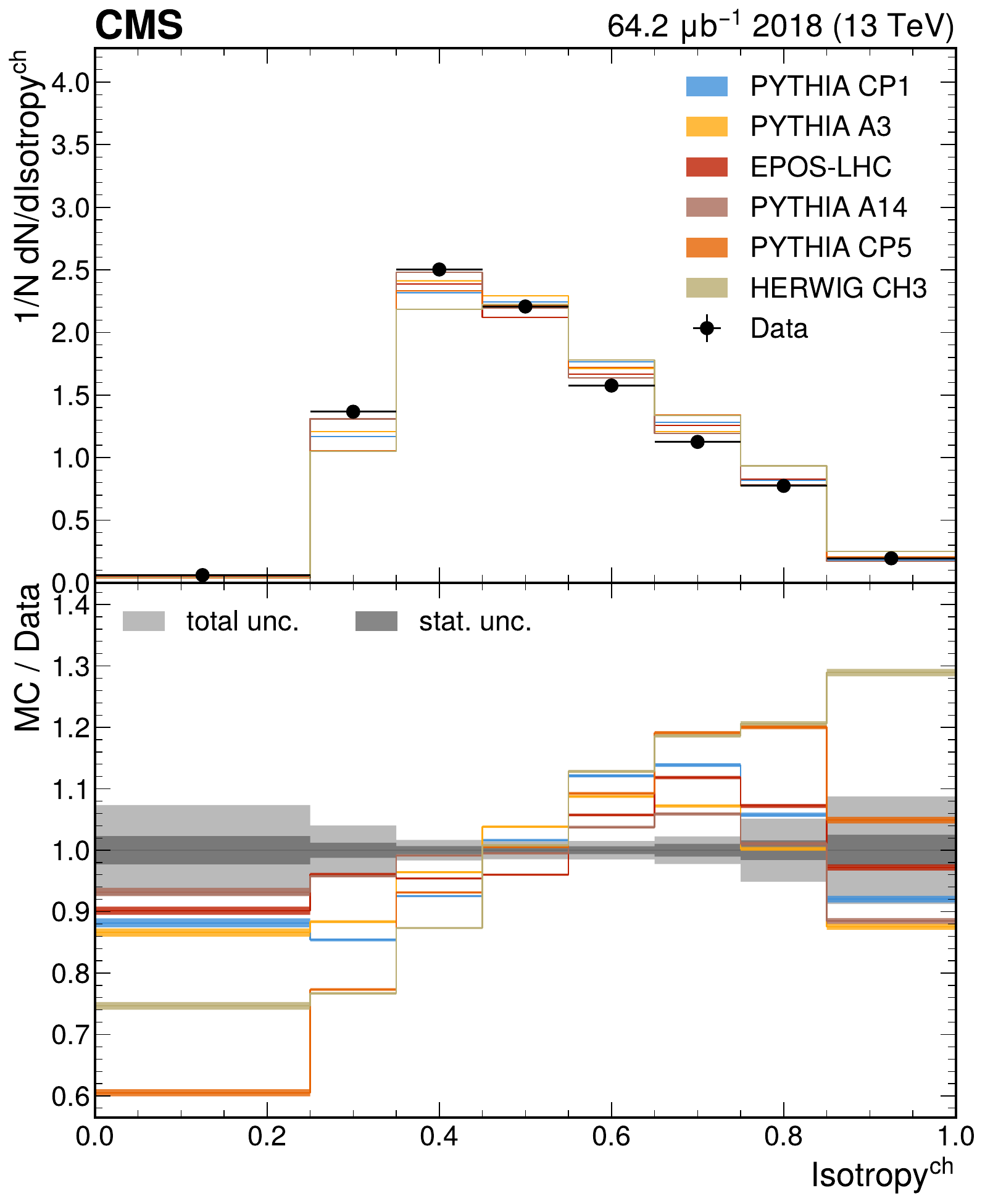}
    \caption{The unfolded results for (upper left) sphericity, (upper right) thrust, (lower left) broadening, and (lower right) isotropy compared with the nominal MC from the \PYTHIA CP1 tune and MC predictions from the \PYTHIA A14, CP5, A3 tunes, the \EPOS generator, and the \HERWIG CH3 tune. The statistical uncertainty on each of the predictions is shown by its shaded band.}
    \label{fig:unfold-gen-1d-2}
\end{figure*}

\begin{figure*}[!ht]
    \centering
    \includegraphics[width=0.48\textwidth]{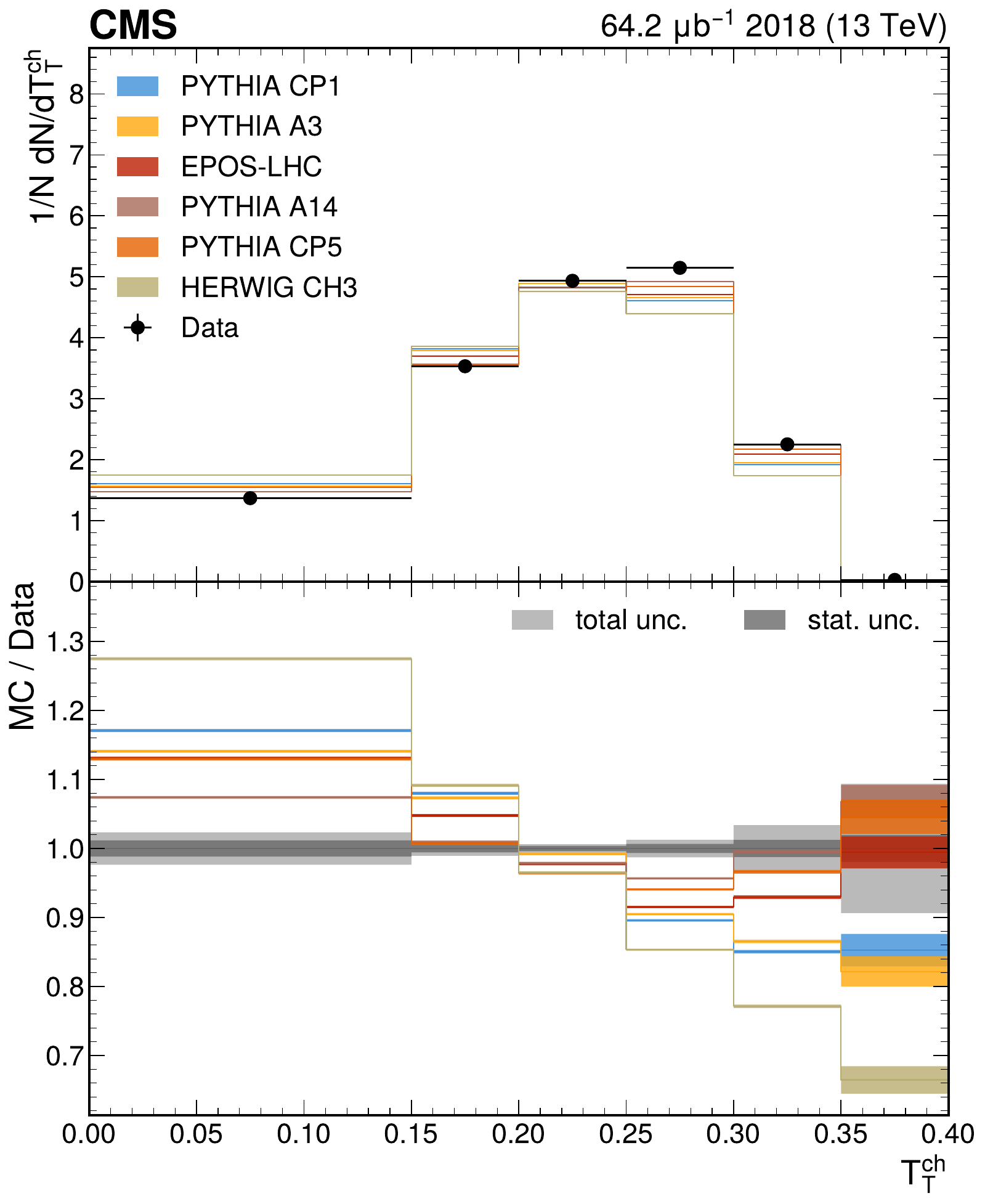}
    \includegraphics[width=0.48\textwidth]{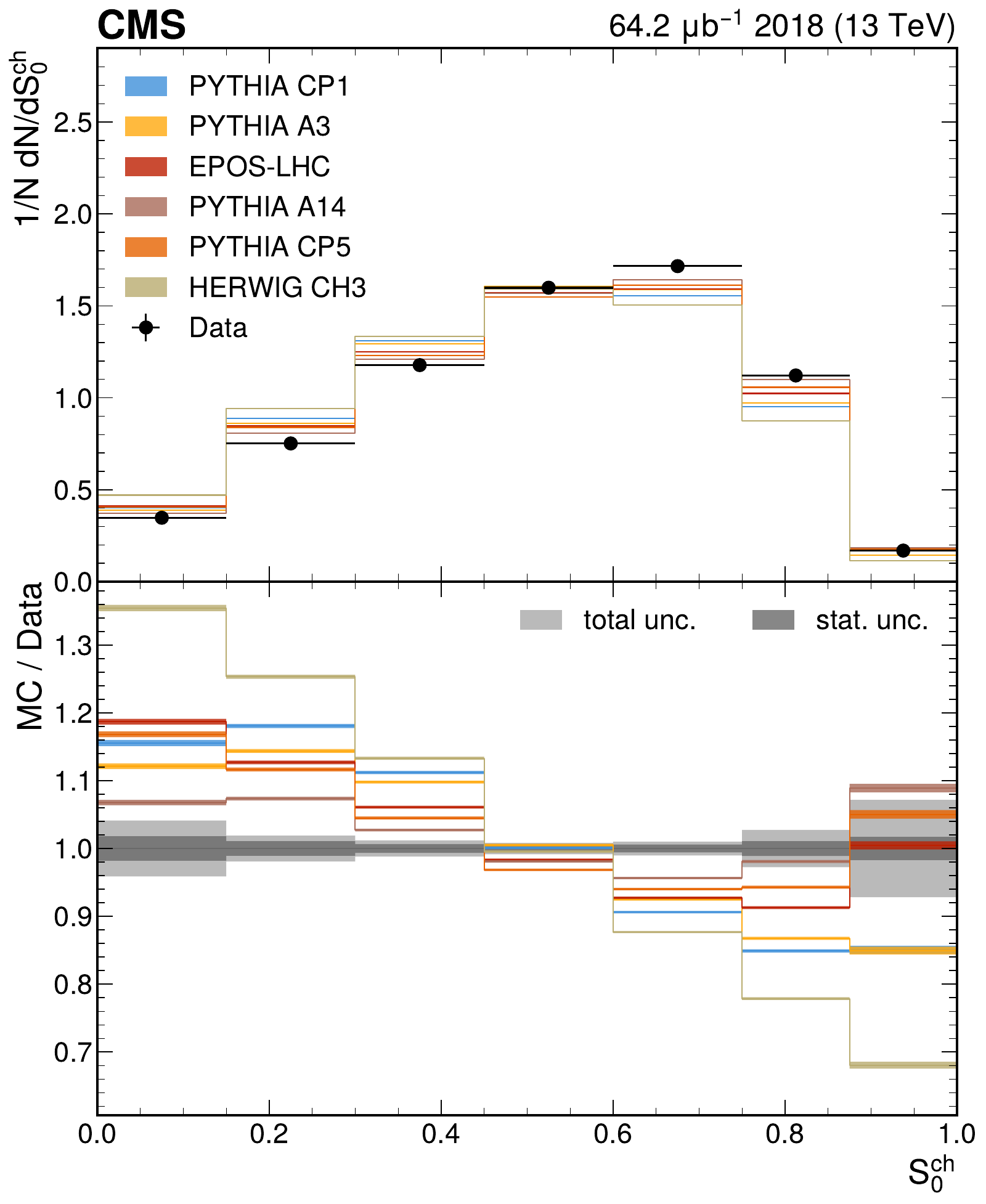}
    \caption{The unfolded results for (left) transverse thrust, and (right) transverse spherocity compared with the nominal MC from the \PYTHIA CP1 tune and predictions from the \PYTHIA A14, CP5, A3 tunes, the \EPOS generator, and the \HERWIG CH3 tune. The statistical uncertainty on each of the predictions is shown by its shaded band. }
    \label{fig:unfold-gen-1d-3}
\end{figure*}

Figures~\ref{fig:unfold-gen-1d-1}--\ref{fig:unfold-gen-1d-3} show the distributions of unfolded observables in 1D using the \PYTHIA CP1 tune as the nominal model. 
    
The number of charged particles and their invariant mass, shown in Fig.~\ref{fig:unfold-gen-1d-1}, are different in data and simulation. The trend of the simulation fluctuates around the data, and does not follow a single monotonic trend. All generators tend to overestimate the relative number of events at low charged-particle multiplicity ($\lesssim 10$), compensated by an underestimate at intermediate values. At high multiplicity and masses, the predictions diverge from each other, and the predictions span the range from underestimating to overestimating the data. The same trend is observed for both the number of charged particles and their invariant mass.
    
The event-shape observables shown in Fig.~\ref{fig:unfold-gen-1d-2} show a consistent trend in which the unfolded data are more isotropic than the simulation. This trend is present for all measures of the event shapes used here, and for all generator configurations considered. Although the level of agreement, particularly in the tails of the distributions, depends on the observables and generator setup being considered, the overall trend is consistent and none of the simulations correctly capture the data.
    
Event-shape observables using only the transverse component of particle momenta are shown in Fig.~\ref{fig:unfold-gen-1d-3}. They show very similar trends to the event-shape observables that include longitudinal momentum information, with the data distributions being more isotropic than the simulations.
  
\begin{figure*}[!ht]
    \centering
    \includegraphics[width=0.9\textwidth]{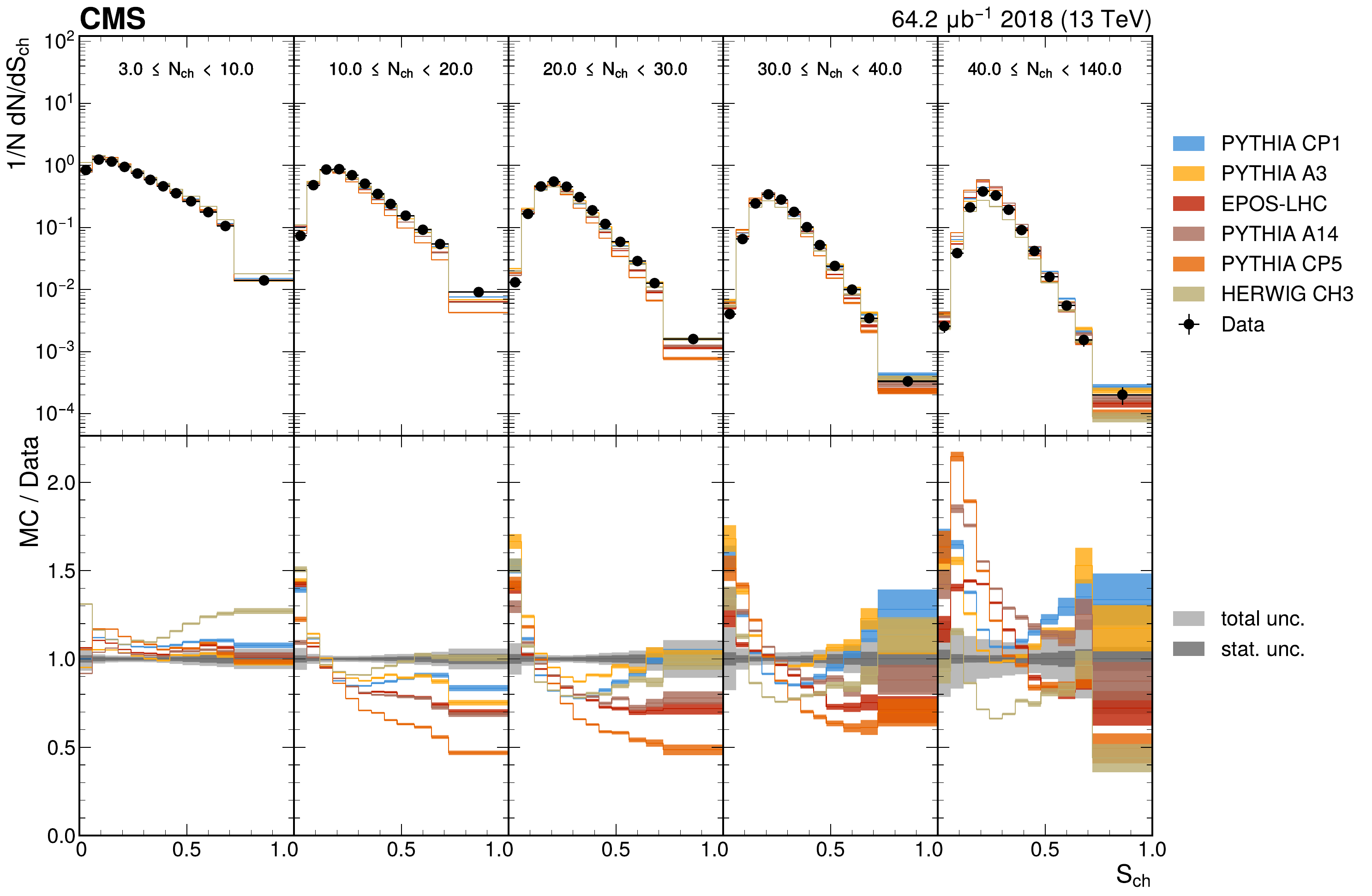}
    \includegraphics[width=0.9\textwidth]{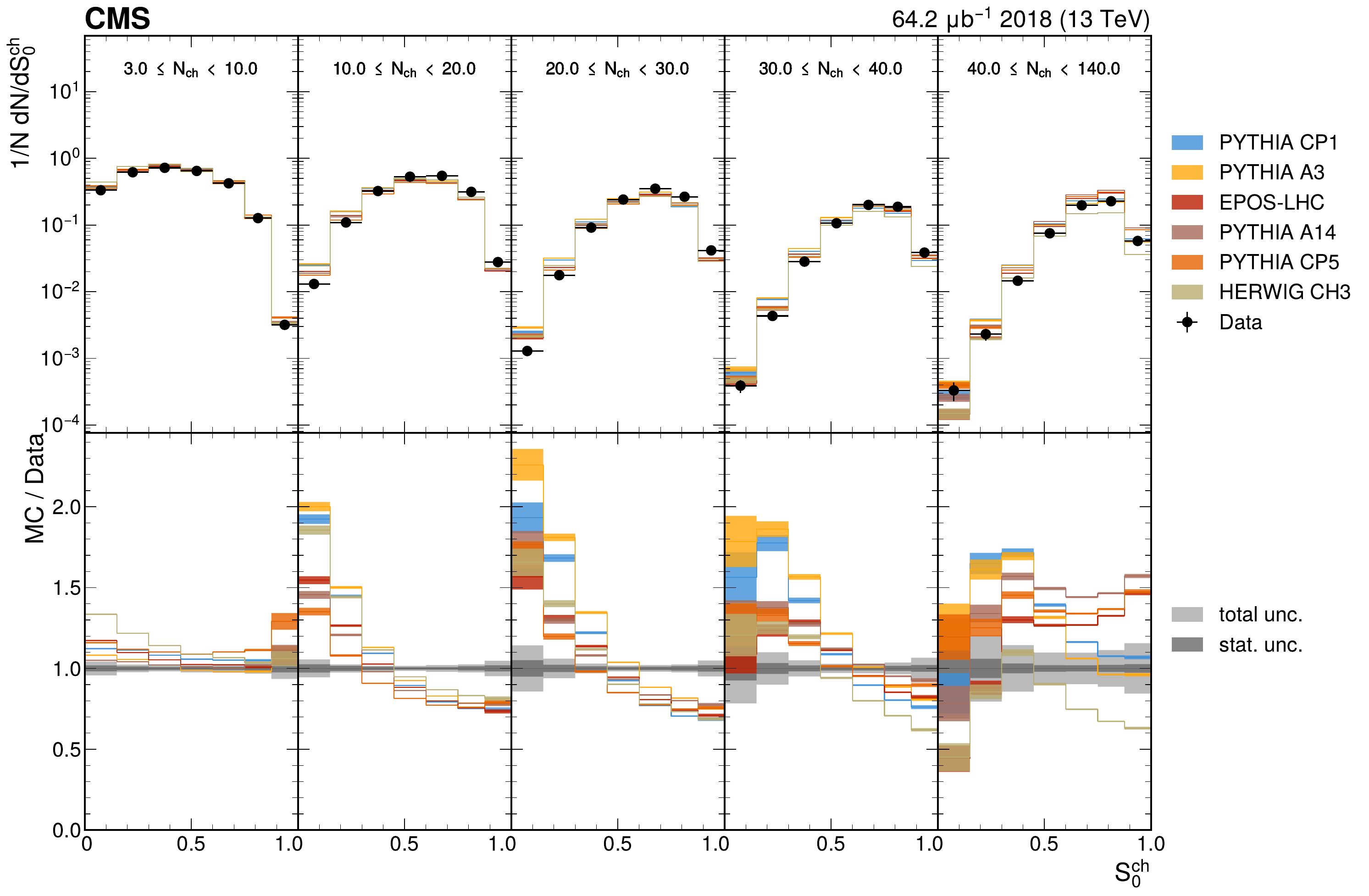}
    \caption{The unfolded distributions of sphericity (upper) and transverse spherocity (lower) in slices of charged-particle multiplicity compared with the nominal MC from the \PYTHIA CP1 tune and MC predictions from the \PYTHIA A14, CP5, A3 tunes, the \EPOS generator, and the \HERWIG CH3 tune. The statistical uncertainty on each of the predictions is shown by its shaded band.}
    \label{fig:unfold-gen-2d}
\end{figure*}

Figure~\ref{fig:unfold-gen-2d} shows the unfolded sphericity and transverse spherocity distributions in slices of the charged-particle multiplicity. Other observables are included in Appendix~\ref{sec:other-2d-distributions}.
    
Within most slices of charged-particle multiplicity, the same general trend of the data being more isotropic than the simulation is present. This occurs over different charged-particle multiplicity slices where the trend of data to simulation in each slice differs, suggesting that the poor modeling of event shapes is likely not primarily coming from the distribution of the number of charged particles. This observation has also been confirmed by looking at slices with a single charged-particle multiplicity, and again is consistent across generator setups. 
    
Checks with the generator setups listed in Table~\ref{table:additional_samples}, including different tunes, PDF variations, and variations of $\alpS$ used in the parton shower, all show the same qualitative behavior.
    
Comparisons to event-shape measurements performed at LEP~\cite{ALEPH:2003obs} were previously performed for the CMS \PYTHIA tunes~\cite{CMS:2019csb}, where very different trends were observed. In those comparisons, which used measurements done in $\Pep\Pem\to\Pq\Paq$ events, the CP2 tune, very similar to the CP1 tune used in this analysis, was found to predict too many isotropic events. The opposite trends of modeling deficiencies for the event shapes of hadrons from \Pep\Pem collisions at LEP and from \pp collisions in this measurement suggest that the poor modeling observed in this analysis is not related to the hadronic final-state evolutions to a large extent, but instead related to the initial states of \pp collisions.
    
Furthermore, the effect of varying $\alpS$ in the final-state shower was checked by using the CP5 $\alpS$ variation tunes~\cite{CMS:2019csb}, which were derived by fitting the $\alpS$ parameter while fixing other values to their CP5 settings, and then taking the variations corresponding to the bounds of the 68\% confidence interval estimated for $\alpS$ from the fit. The predictions for the measured distributions have only minor changes, less than what is observed for other variations of settings shown here. The CUEP8M1~\cite{CMS:2015wcf} and CUEP8M2T4~\cite{CMS-PAS-TOP-16-021,Sirunyan_2019} tunes both gave results very similar to those for the CP1 tune. The \PYTHIA 6 Z2Star tune~\cite{CMS:2011qzf} shows larger discrepancies with the measured distributions than the models considered in the figures, probably because it was based on measurements of \pp collisions at energies lower than 13\TeV.
    
These results are consistent with what was observed in a study of underlying event modeling in \ttbar events at CMS~\cite{Sirunyan_2019}, where the underlying event simulations were observed to be somewhat less isotropic than the recorded data.

\section{Summary}
A measurement of event shapes in a minimum-bias selection of proton-proton collisions at a center-of-mass energy of 13\TeV has been presented. Low-pileup data collected with the CMS detector in 2018 were used, and the kinematics of reconstructed tracks were used to unfold the distributions to the level of stable charged particles.
An unbinned multidimensional unfolding algorithm was used to obtain these results. All one- and two-dimensional distributions as well as their covariances are provided.

The results show a consistent trend of mismodeling event shapes common across all simulation configurations considered. The simulations under study include several different underlying physics models, from the \PYTHIA, \EPOS, and \HERWIG generators, as well as a number of tunes of the \PYTHIA generator. For each of these simulations, the isotropy of the event shapes is underpredicted across a range of six different event-shape observables.

Some mismodeling of the charged-particle multiplicity and the invariant mass of the charged particles is also observed. The interplay between these two trends was investigated by considering the event-shape observables in slices of charged-particle multiplicity. These investigations show that the observed data continue to prefer more isotropic distributions even within slices of charged-particle multiplicity and when considering only the plane transverse to the beam direction. These observations suggest that the mismodeling of event shapes is likely not coming only from the distribution of the number of charged particles or poorly modeled longitudinal components.

These event-shape observables are important in probing soft and nonperturbative effects in quantum chromodynamics (QCD) at the LHC. The unfolded data~\cite{hepdata,CDSdata} should be used by the community to improve and develop existing proton-proton collision models. The continued development of these models will be critical for understanding phenomena such as quark-gluon plasma and topological effects in non-Abelian gauge theories, such as QCD instantons.

\begin{acknowledgments}
We congratulate our colleagues in the CERN accelerator departments for the excellent performance of the LHC and thank the technical and administrative staffs at CERN and at other CMS institutes for their contributions to the success of the CMS effort. In addition, we gratefully acknowledge the computing centres and personnel of the Worldwide LHC Computing Grid and other centres for delivering so effectively the computing infrastructure essential to our analyses. Finally, we acknowledge the enduring support for the construction and operation of the LHC, the CMS detector, and the supporting computing infrastructure provided by the following funding agencies: SC (Armenia), BMBWF and FWF (Austria); FNRS and FWO (Belgium); CNPq, CAPES, FAPERJ, FAPERGS, and FAPESP (Brazil); MES and BNSF (Bulgaria); CERN; CAS, MoST, and NSFC (China); MINCIENCIAS (Colombia); MSES and CSF (Croatia); RIF (Cyprus); SENESCYT (Ecuador); ERC PRG, RVTT3 and MoER TK202 (Estonia); Academy of Finland, MEC, and HIP (Finland); CEA and CNRS/IN2P3 (France); SRNSF (Georgia); BMBF, DFG, and HGF (Germany); GSRI (Greece); NKFIH (Hungary); DAE and DST (India); IPM (Iran); SFI (Ireland); INFN (Italy); MSIP and NRF (Republic of Korea); MES (Latvia); LMTLT (Lithuania); MOE and UM (Malaysia); BUAP, CINVESTAV, CONACYT, LNS, SEP, and UASLP-FAI (Mexico); MOS (Montenegro); MBIE (New Zealand); PAEC (Pakistan); MES and NSC (Poland); FCT (Portugal); MESTD (Serbia); MICIU/AEI and PCTI (Spain); MOSTR (Sri Lanka); Swiss Funding Agencies (Switzerland); MST (Taipei); MHESI and NSTDA (Thailand); TUBITAK and TENMAK (Turkey); NASU (Ukraine); STFC (United Kingdom); DOE and NSF (USA).

\hyphenation{Rachada-pisek} Individuals have received support from the Marie-Curie programme and the European Research Council and Horizon 2020 Grant, contract Nos.\ 675440, 724704, 752730, 758316, 765710, 824093, 101115353, 101002207, 101001205, and COST Action CA16108 (European Union); the Leventis Foundation; the Alfred P.\ Sloan Foundation; the Alexander von Humboldt Foundation; the Science Committee, project no. 22rl-037 (Armenia); the Fonds pour la Formation \`a la Recherche dans l'Industrie et dans l'Agriculture (FRIA-Belgium); the Beijing Municipal Science \& Technology Commission, No. Z191100007219010, the Fundamental Research Funds for the Central Universities, the Ministry of Science and Technology of China under Grant No. 2023YFA1605804, and the Natural Science Foundation of China under Grant No. 12061141002 (China); the Ministry of Education, Youth and Sports (MEYS) of the Czech Republic; the Shota Rustaveli National Science Foundation, grant FR-22-985 (Georgia); the Deutsche Forschungsgemeinschaft (DFG), among others, under Germany's Excellence Strategy -- EXC 2121 ``Quantum Universe" -- 390833306, and under project number 400140256 - GRK2497; the Hellenic Foundation for Research and Innovation (HFRI), Project Number 2288 (Greece); the Hungarian Academy of Sciences, the New National Excellence Program - \'UNKP, the NKFIH research grants K 131991, K 133046, K 138136, K 143460, K 143477, K 146913, K 146914, K 147048, 2020-2.2.1-ED-2021-00181, TKP2021-NKTA-64, and 2021-4.1.2-NEMZ\_KI-2024-00036 (Hungary); the Council of Science and Industrial Research, India; ICSC -- National Research Centre for High Performance Computing, Big Data and Quantum Computing and FAIR -- Future Artificial Intelligence Research, funded by the NextGenerationEU program (Italy); the Latvian Council of Science; the Ministry of Education and Science, project no. 2022/WK/14, and the National Science Center, contracts Opus 2021/41/B/ST2/01369 and 2021/43/B/ST2/01552 (Poland); the Funda\c{c}\~ao para a Ci\^encia e a Tecnologia, grant CEECIND/01334/2018 (Portugal); the National Priorities Research Program by Qatar National Research Fund; MICIU/AEI/10.13039/501100011033, ERDF/EU, "European Union NextGenerationEU/PRTR", and Programa Severo Ochoa del Principado de Asturias (Spain); the Chulalongkorn Academic into Its 2nd Century Project Advancement Project, and the National Science, Research and Innovation Fund via the Program Management Unit for Human Resources \& Institutional Development, Research and Innovation, grant B39G670016 (Thailand); the Kavli Foundation; the Nvidia Corporation; the SuperMicro Corporation; the Welch Foundation, contract C-1845; and the Weston Havens Foundation (USA).
\end{acknowledgments}

\bibliography{auto_generated} 

\appendix
\numberwithin{figure}{section}
\numberwithin{table}{section}

\section{Additional simulated samples for validation}\label{sec:additional_samples}
Table~\ref{table:additional_samples} gives additional simulated samples that are used as pseudodata to validate the performance of the unfolding algorithm or to extensively investigate the impacts of various models on the observables to be measured.
\begin{table*}[htbp!]
    \centering
    \topcaption{Samples of simulated minimum-bias events used for validation. The value of $\alpS$ corresponds to that used in the PDF set. }
    \cmsTable{
    \begin{scotch}{lll}
      & Generator & PDF  ($\alpS(m_{\PZ}) $) \\
     \hline
     Pseudodata samples & \PYTHIA{8} (CP5)~\cite{CMS:2019csb} & NNPDF3.1 QCD+LUXQED NNLO  (0.118)  \\
                                         &   \PYTHIA{8} (CUETP8M1)~\cite{CMS:2015wcf} & NNPDF2.3 QCD+QED LO (0.130) \\[\cmsTabSkip]
     Other validation & \PYTHIA{8} (CP5) with pileup & NNPDF3.1 QCD+LUXQED NNLO  (0.118)  \\
                                     & \PYTHIA{8} (CP5) $\alpS^{\text{FSR}}$ variations~\cite{CMS:2019csb} & NNPDF3.1 QCD+LUXQED NNLO  (0.118)  \\         
                                     & \PYTHIA{8} CUETP8M2T4~\cite{CMS-PAS-TOP-16-021,Sirunyan_2019} & NNPDF3.0 QCD LO (0.130)  \\
                                     & \PYTHIA{8} (CP5) color reconnection tunes~\cite{CMS:2022awf} & NNPDF3.1 QCD+LUXQED NNLO (0.118)  \\
                                     & \PYTHIA{8} (CP2) & NNPDF3.1 QCD LO (0.130)  \\
                                     & \PYTHIA{8} (A14)~\cite{ATL-PHYS-PUB-2014-021} & NNPDF3.1 QCD LO  (0.130)  \\
                                     & \PYTHIA{8} (A14) eigenvariations & NNPDF3.1 QCD LO  (0.130)  \\
                                     & \PYTHIA{8} (A14) CTEQL1 & CTEQL1  (0.1298)  \\
                                     & \PYTHIA{8} (A14) MSTW2008LO & MSTW2008LO  (0.13939)  \\
                                     & \PYTHIA{8} (A14) HERAPDF1.5LO & HERAPDF1.5LO  (0.130)  \\
    \end{scotch}}
    \label{table:additional_samples}
\end{table*}

\section{Additional two-dimensional distributions}\label{sec:other-2d-distributions}
\begin{figure*}[!ht]
\centering
\includegraphics[width=0.95\textwidth]{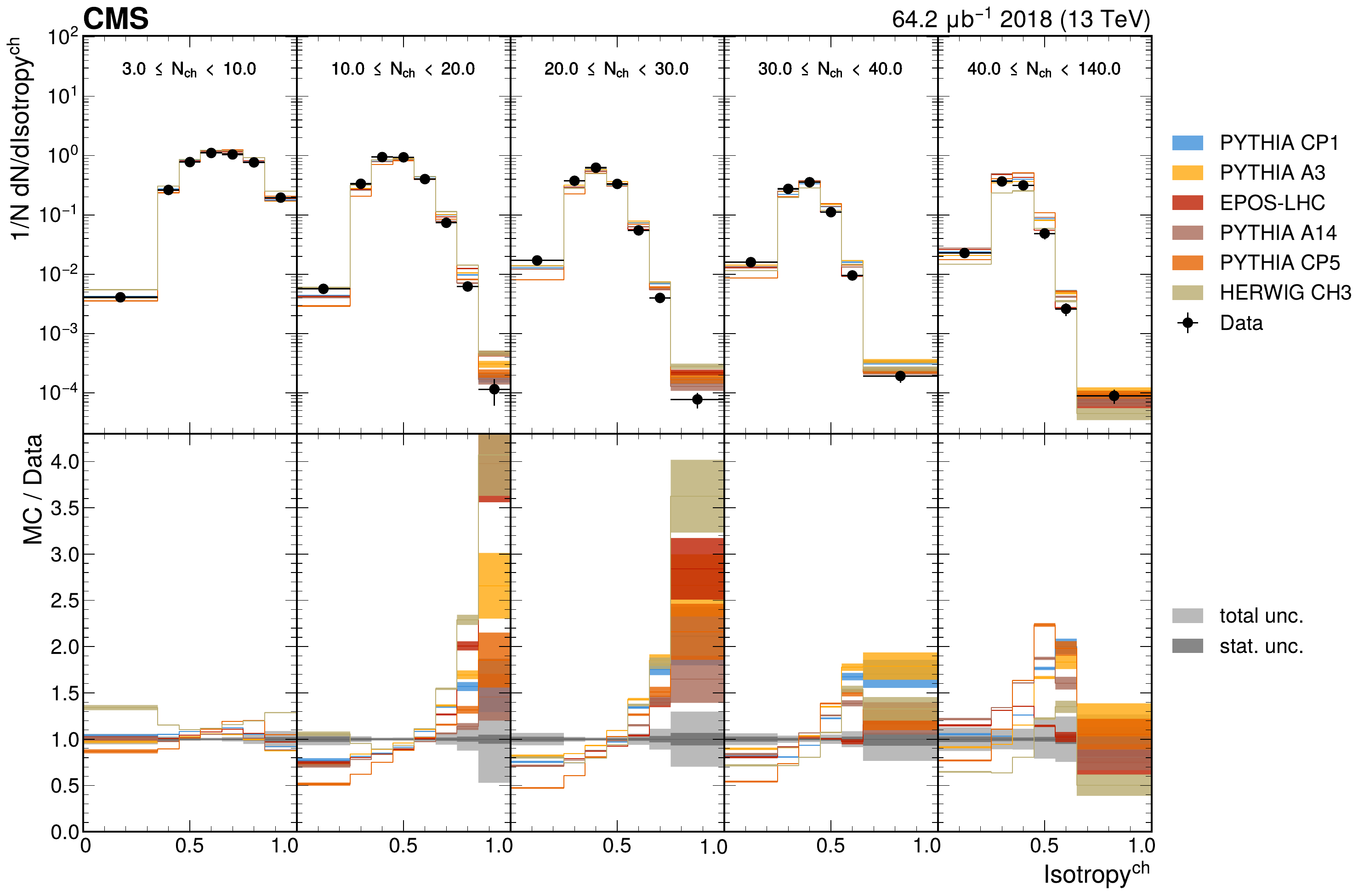}
\includegraphics[width=0.95\textwidth]{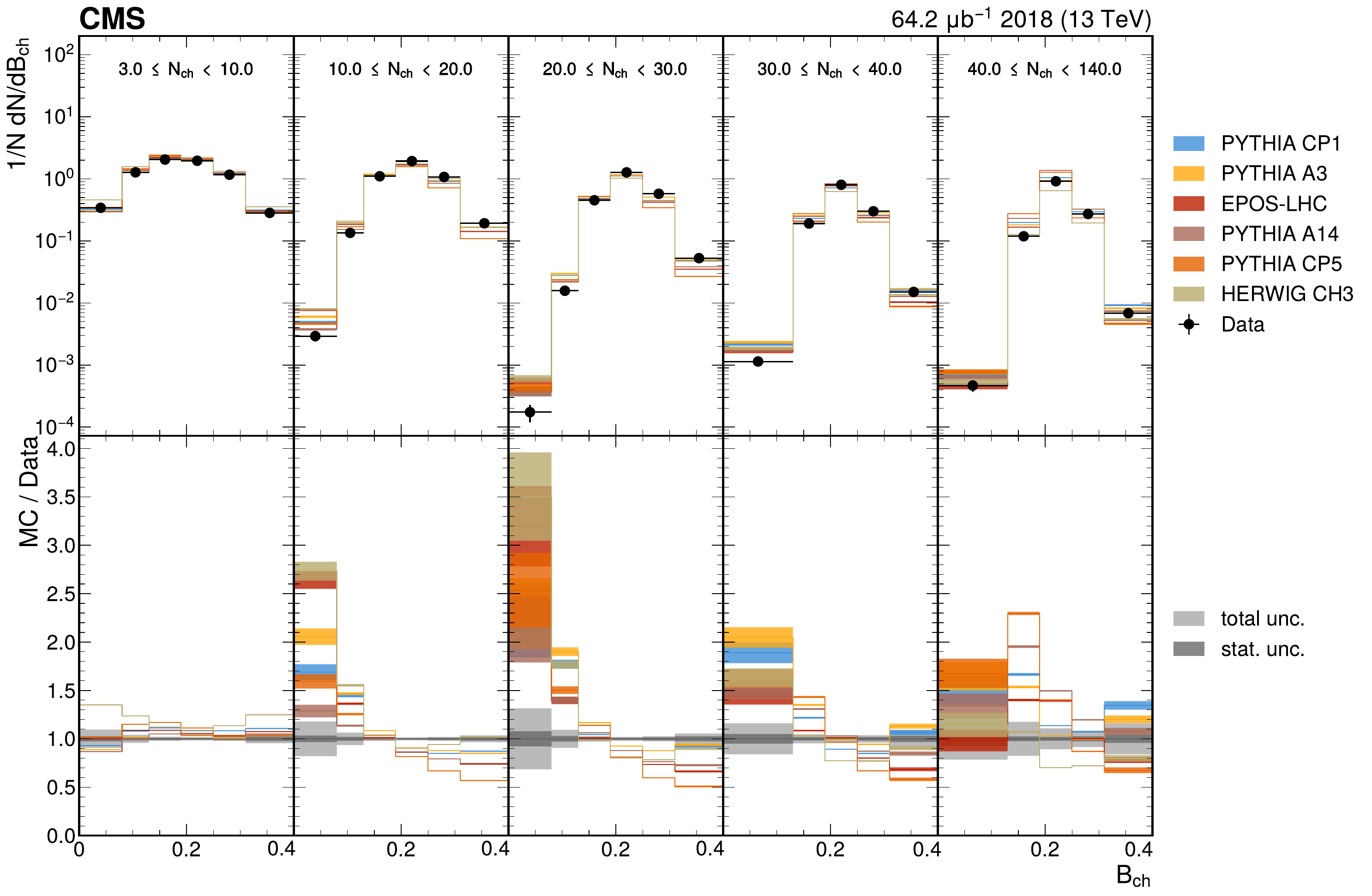}
\caption{The unfolded distributions of event isotropy (upper) and broadening (lower) in slices of charged-particle multiplicity compared with the nominal MC from the \PYTHIA CP1 tune and MC predictions from the \PYTHIA A14, CP5, A3 tunes, the \EPOS generator, and the \HERWIG CH3 tune. The statistical uncertainty on each of the predictions is shown by its shaded band.}
\label{fig:unfold-gen-2db}
\end{figure*}

\begin{figure*}[!ht]
\centering
\includegraphics[width=0.95\textwidth]{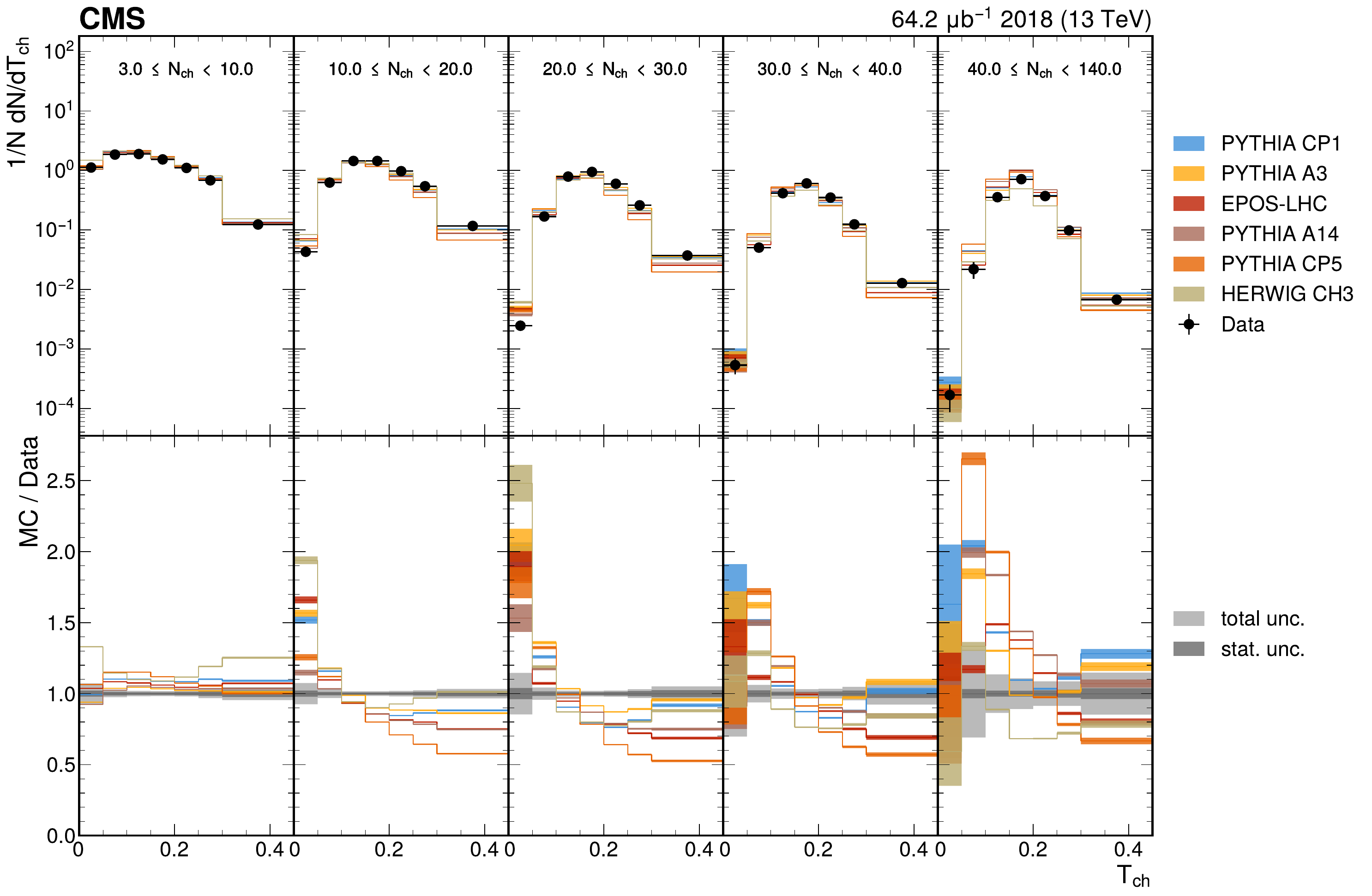}
\includegraphics[width=0.95\textwidth]{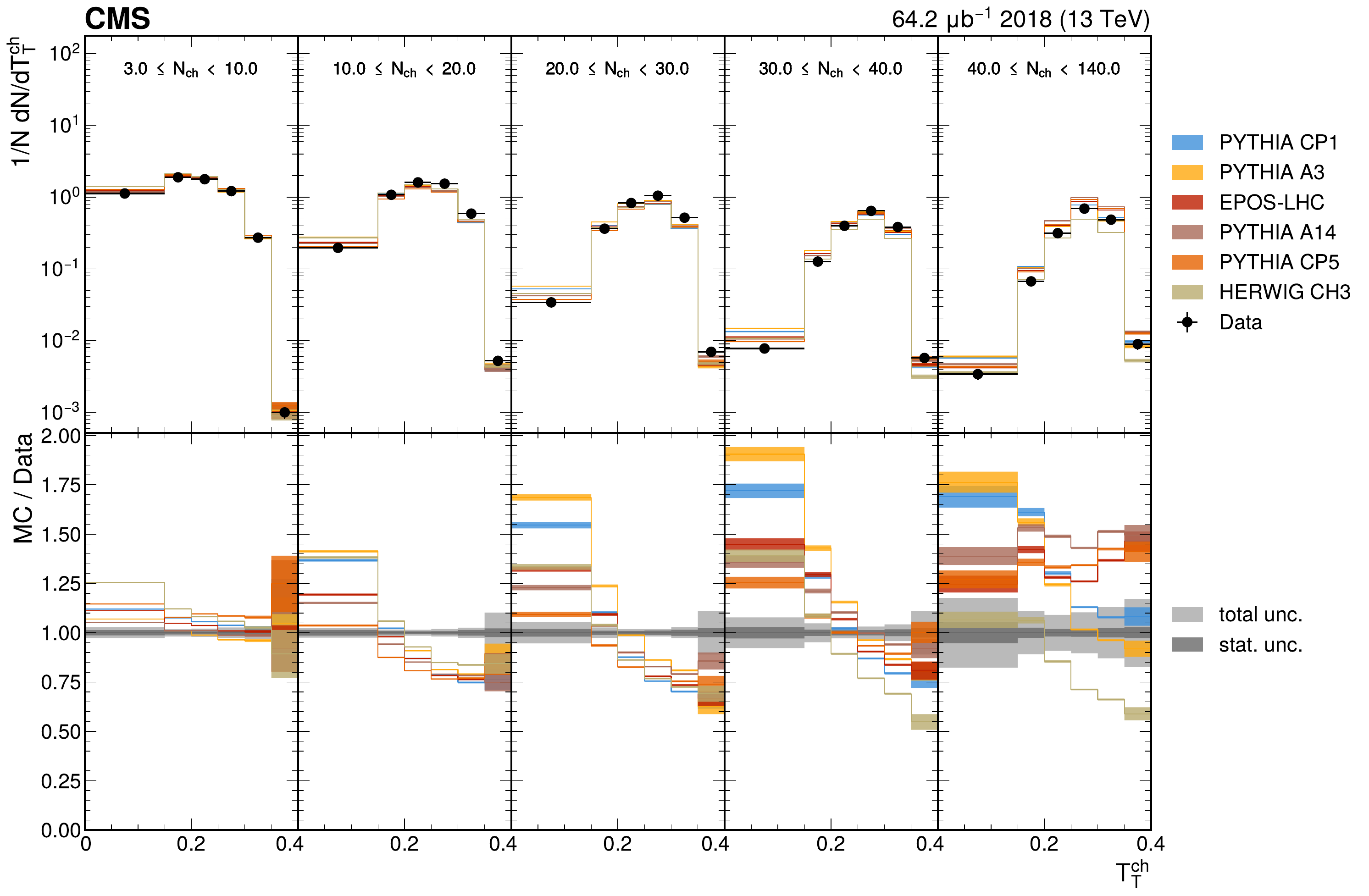}
\caption{The unfolded distributions of thrust (upper) and transverse thrust (lower) in slices of charged-particle multiplicity compared with the nominal MC from the \PYTHIA CP1 tune and MC predictions from the \PYTHIA A14, CP5, A3 tunes, the \EPOS generator, and the \HERWIG CH3 tune. The statistical uncertainty on each of the predictions is shown by its shaded band.}
\label{fig:unfold-gen-2dc}
\end{figure*}

\begin{figure*}[!ht]
\centering
\includegraphics[width=0.95\textwidth]{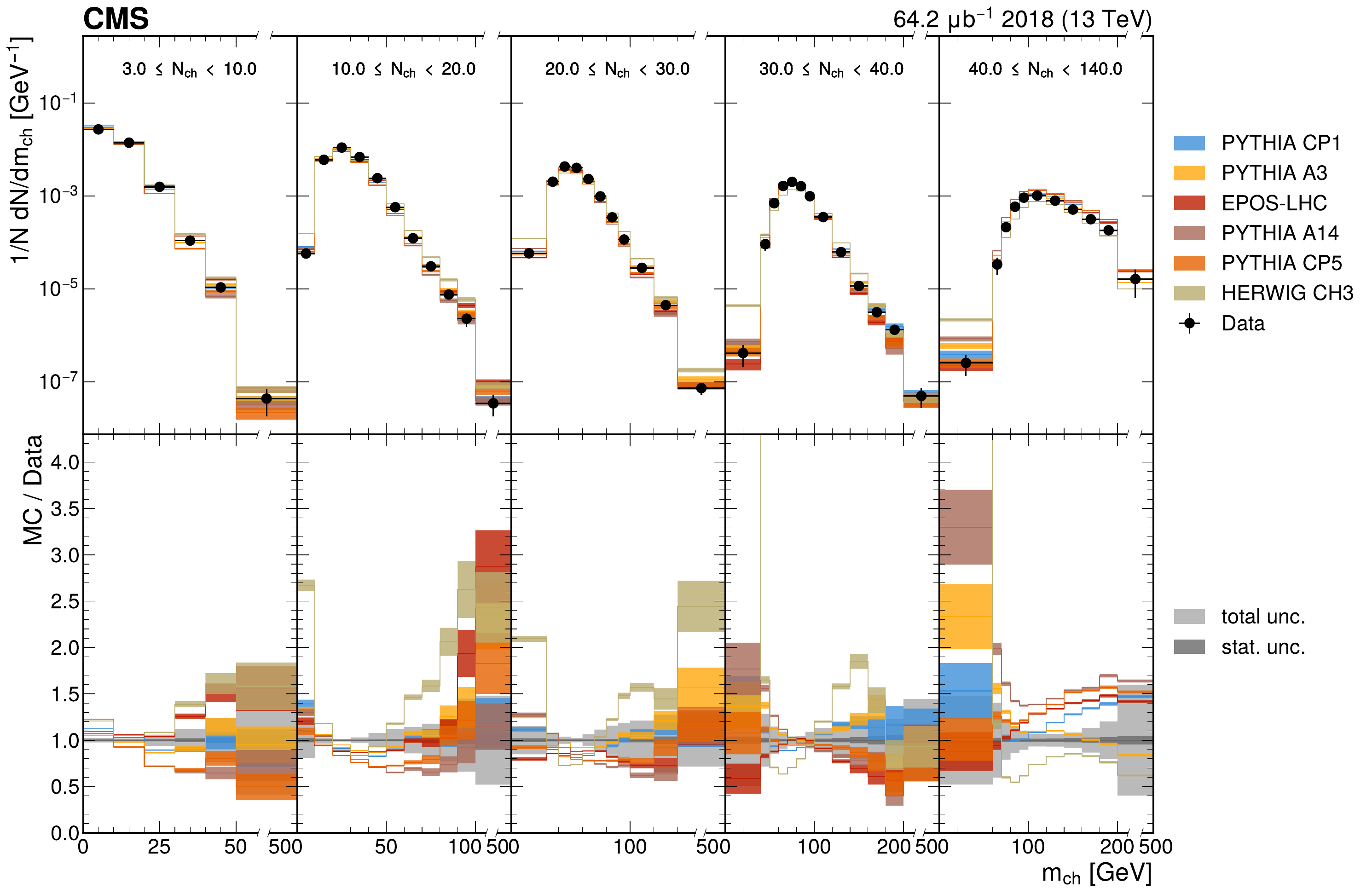}
\caption{The unfolded invariant mass of charged particles in slices of charged-particle multiplicity compared with the nominal MC from the \PYTHIA CP1 tune and MC predictions from the \PYTHIA A14, CP5, A3 tunes, the \EPOS generator, and the \HERWIG CH3 tune. The statistical uncertainty on each of the predictions is shown by its shaded band.}
\label{fig:unfold-gen-2dd}
\end{figure*}

Figures~\ref{fig:unfold-gen-2db}, \ref{fig:unfold-gen-2dc} and \ref{fig:unfold-gen-2dd} show the unfolded distributions of isotropy, broadening, thrust, transverse thrust and invariant mass of charged particles in slices of the charged-particle multiplicity, as well as the MC predictions for comparison.
\clearpage

\section{Correlation matrices}\label{sec:correlation}
\begin{figure*}[!ht]
\centering
\includegraphics[width=0.45\textwidth]{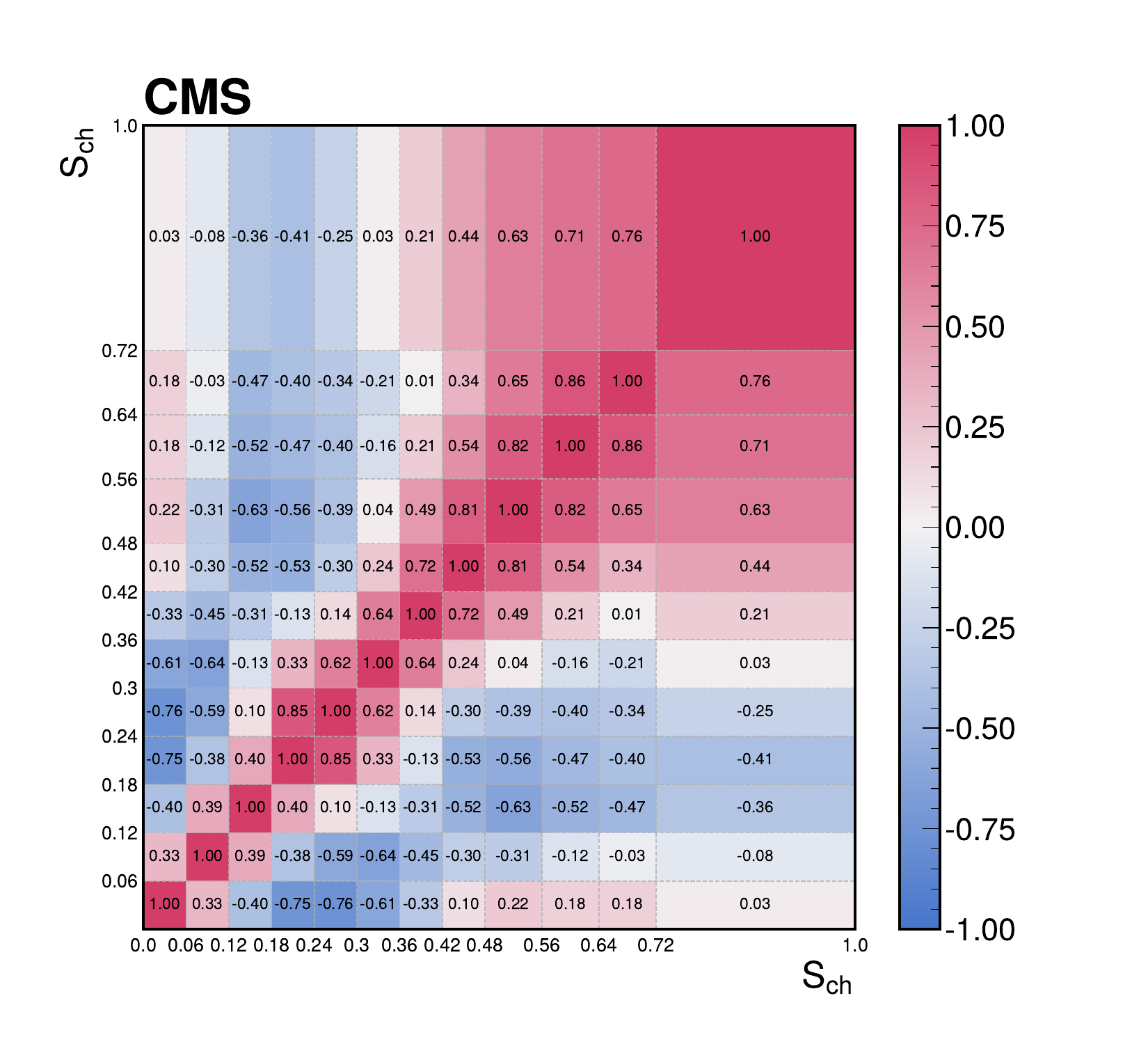}
\includegraphics[width=0.45\textwidth]{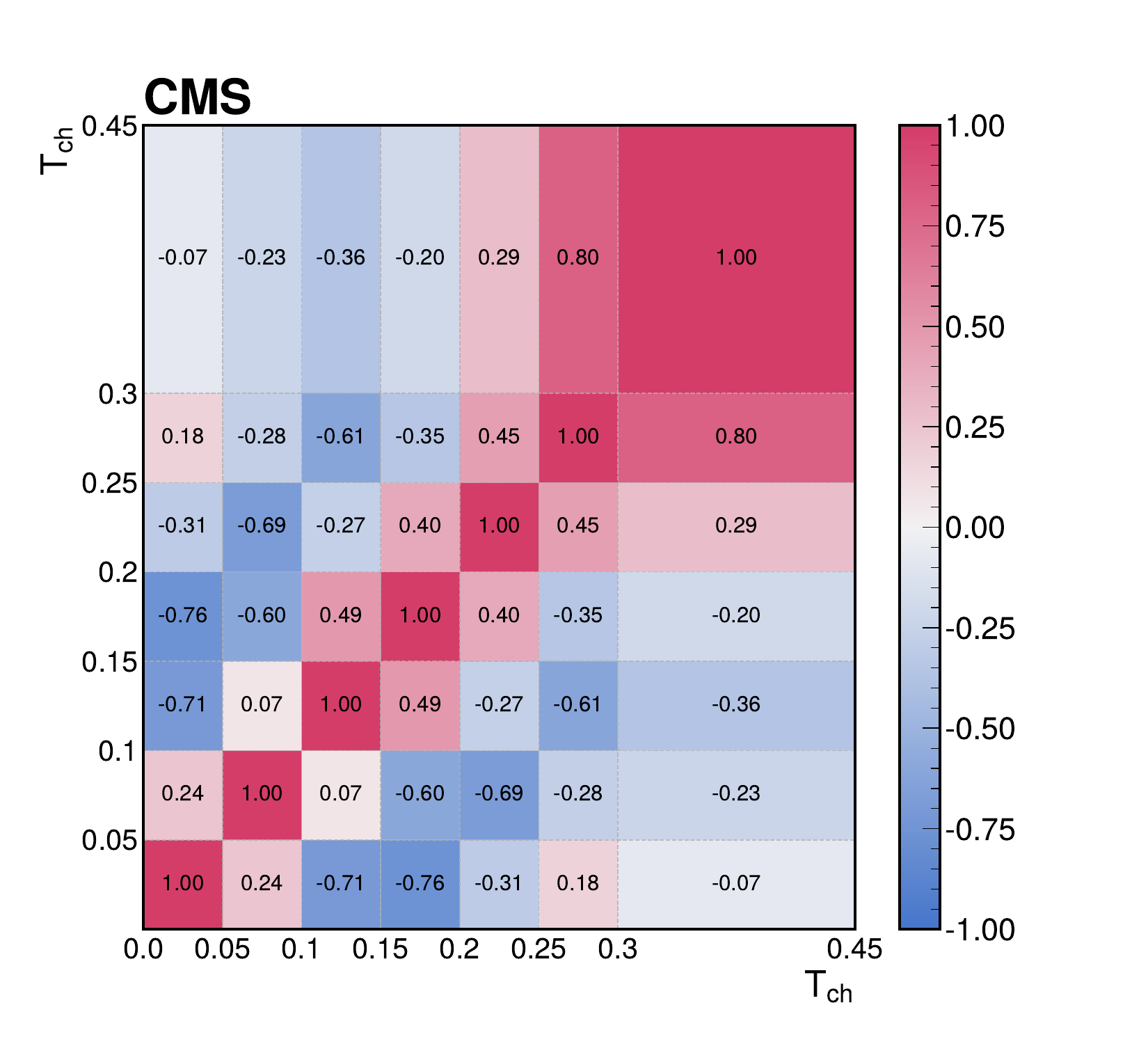}
\includegraphics[width=0.45\textwidth]{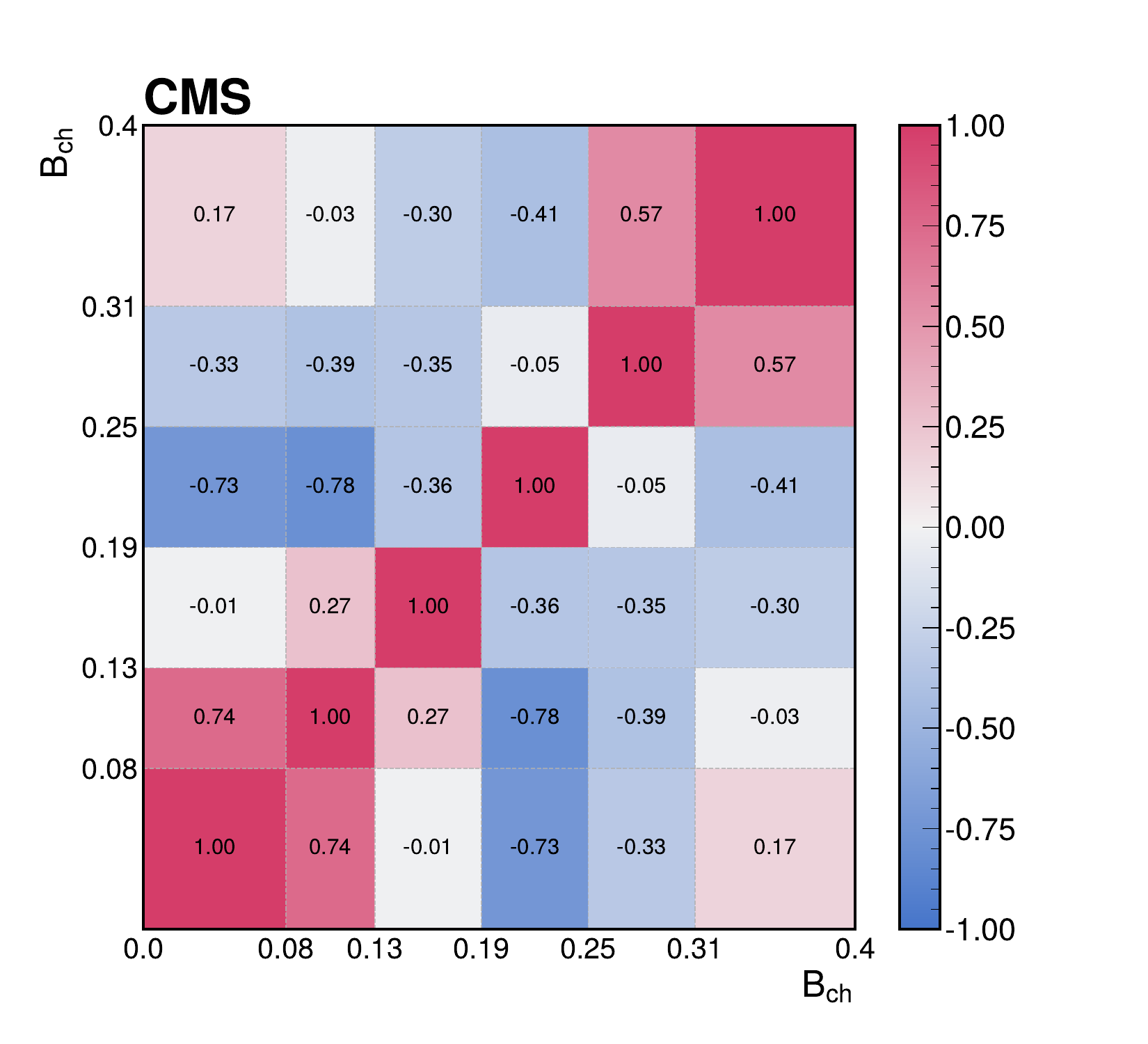}
\includegraphics[width=0.45\textwidth]{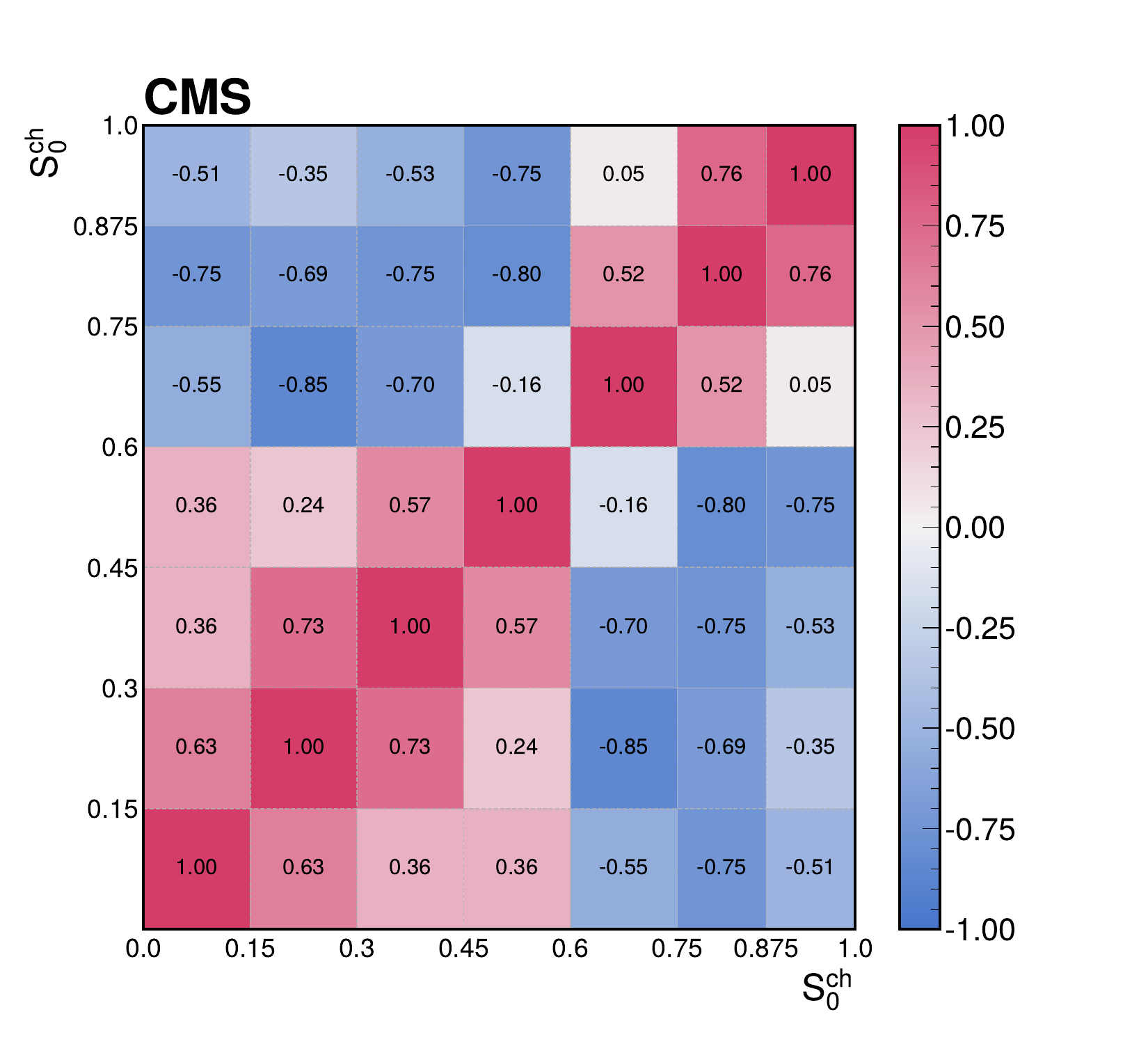}
\caption{The correlations of the unfolding systematic uncertainty between bins of the event-shape observables shown for (upper left) sphericity, (upper right) thrust, (lower left) broadening, and (lower right) transverse spherocity.}
\label{fig:unfold-corrsys-1d-3}
\end{figure*}

\begin{figure*}[!ht]
\centering
\includegraphics[width=0.45\textwidth]{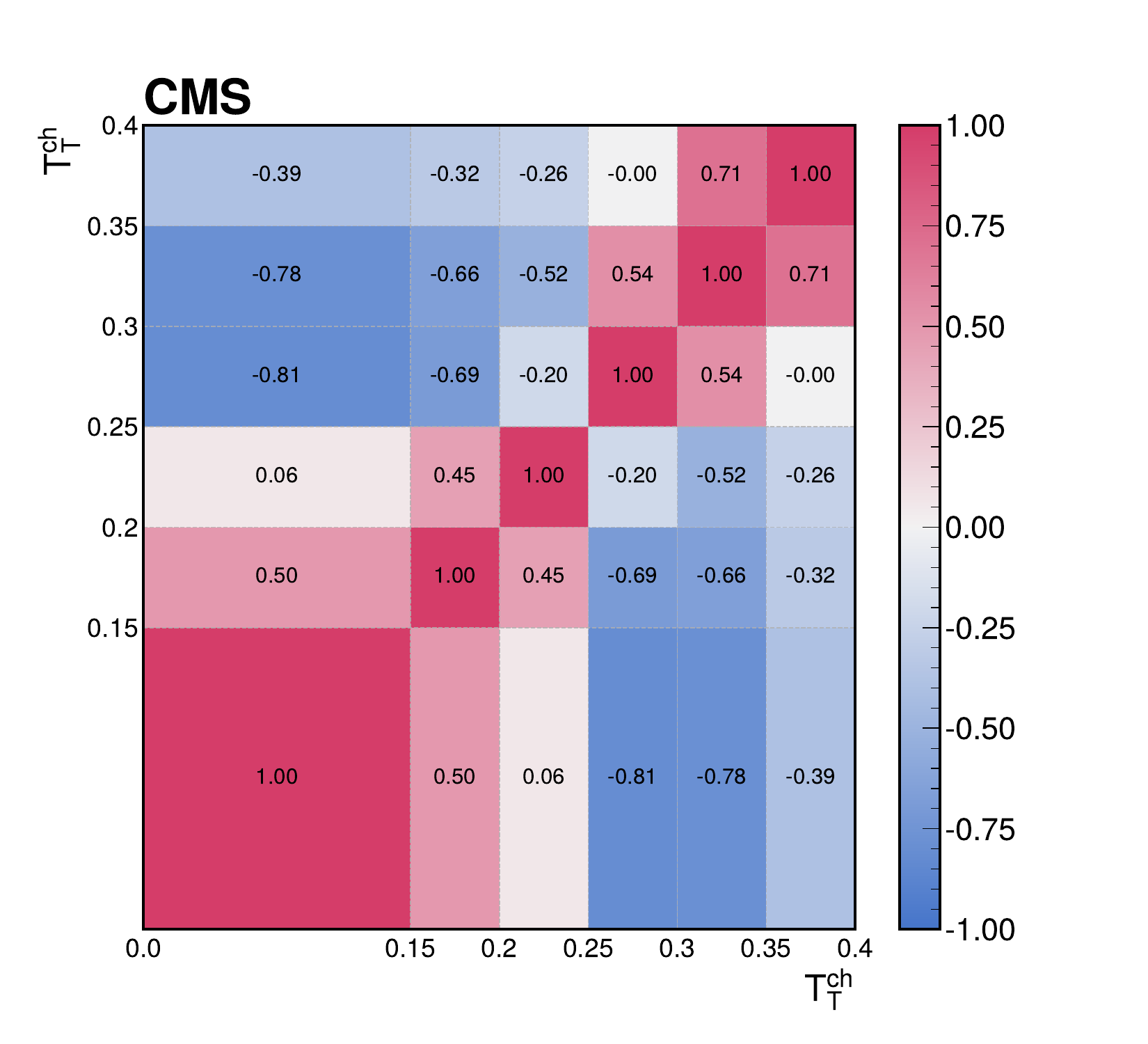}
\includegraphics[width=0.45\textwidth]{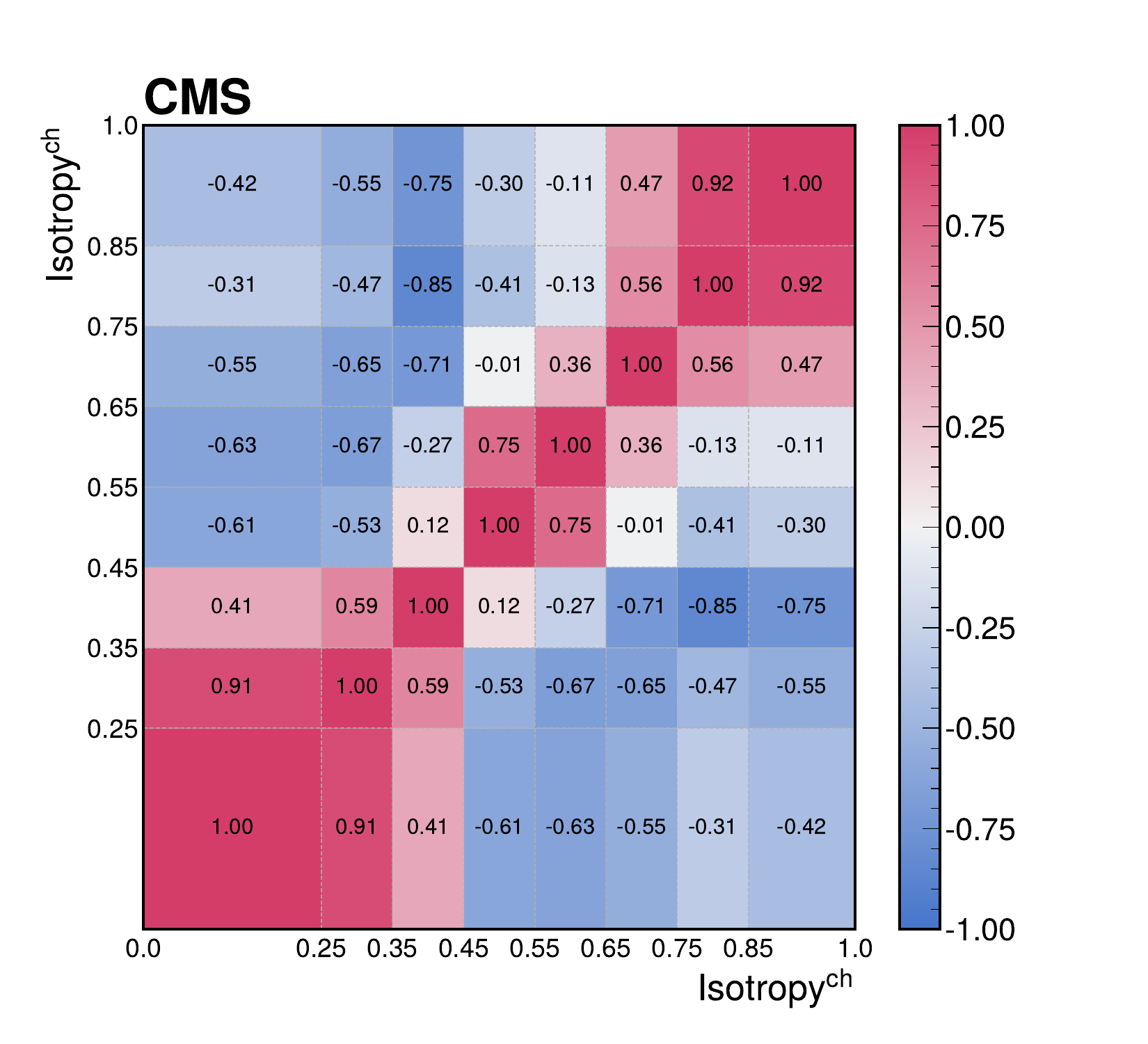}
\includegraphics[width=0.45\textwidth]{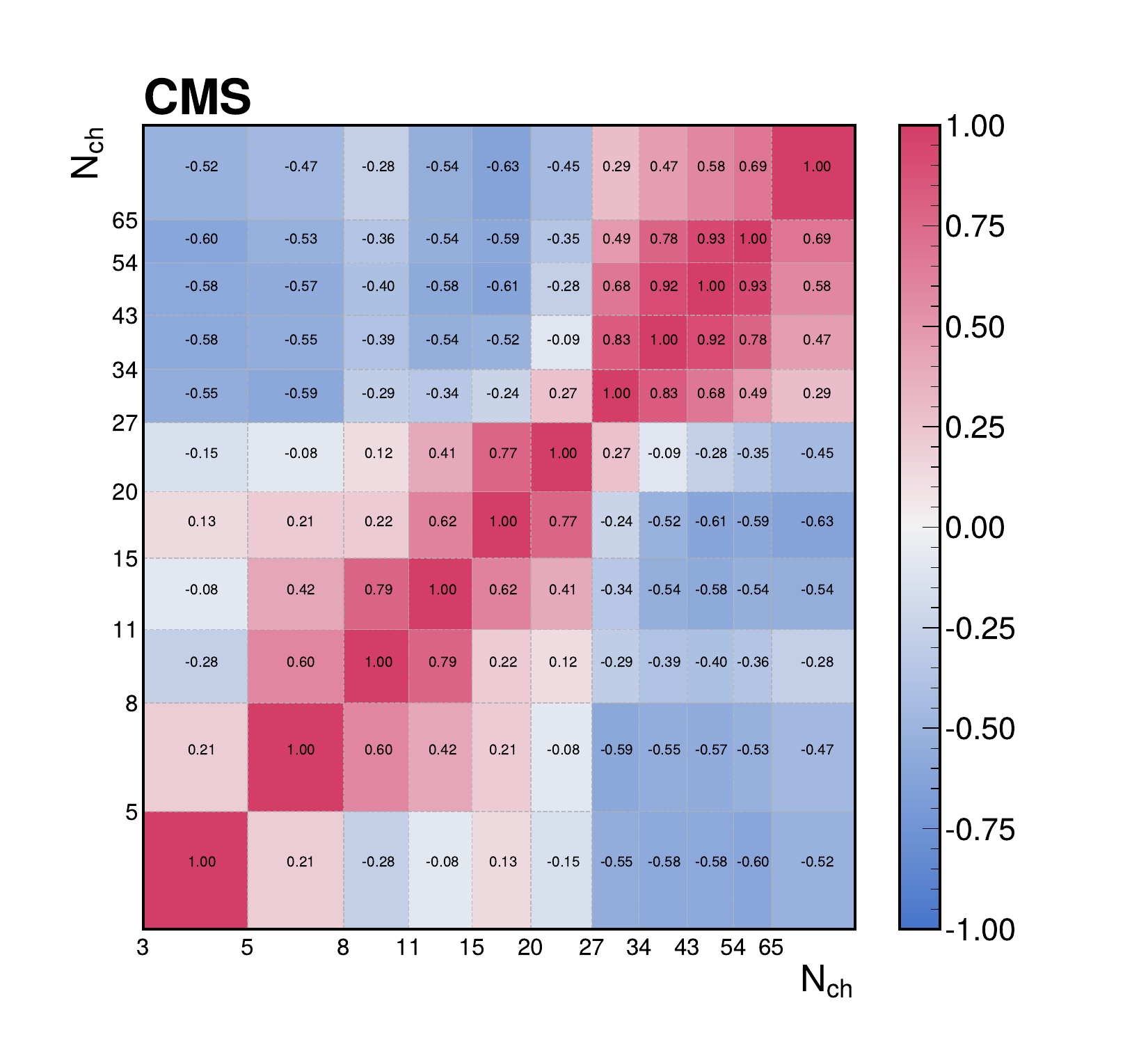}
\includegraphics[width=0.45\textwidth]{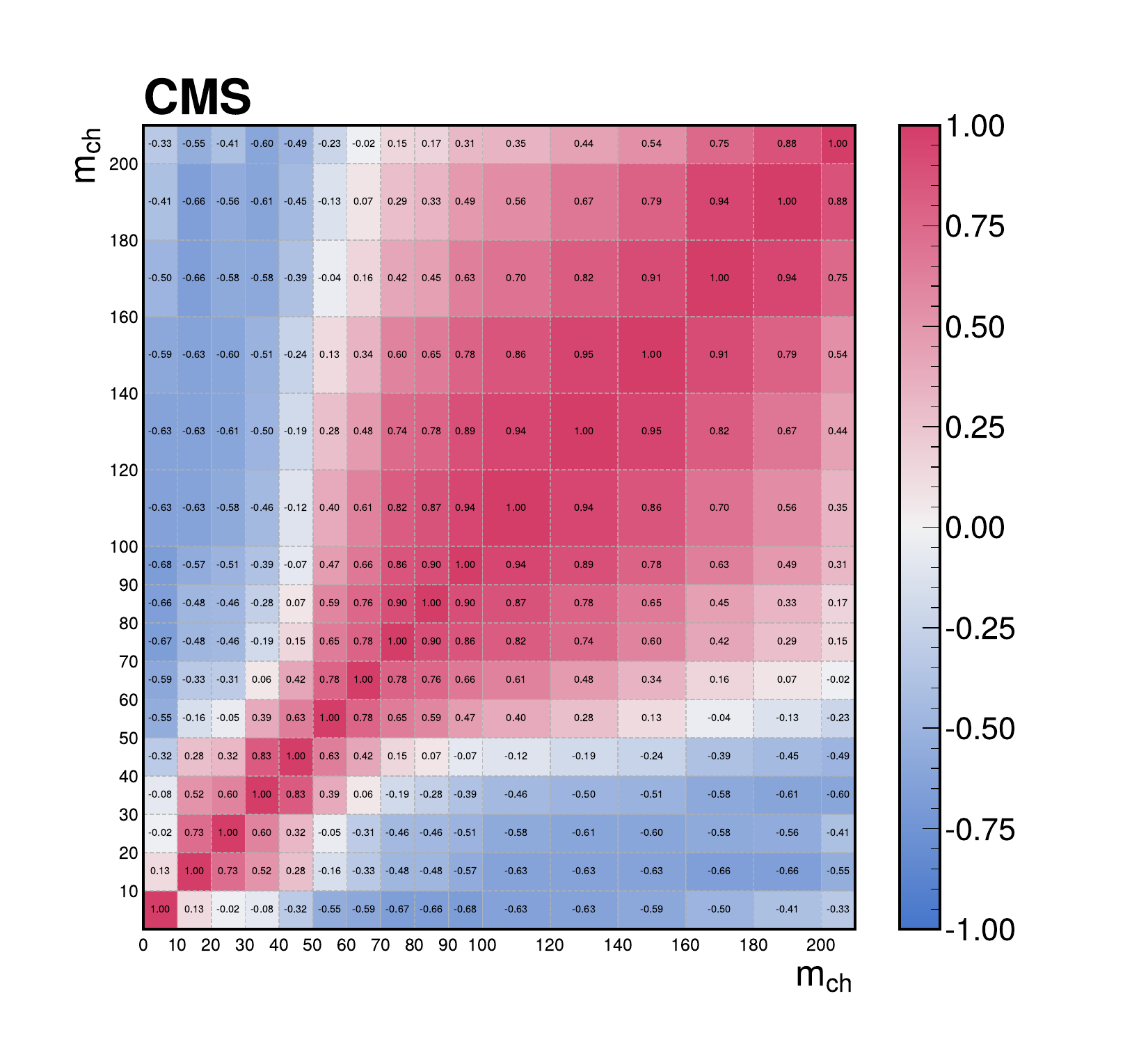}
\caption{The correlations of the unfolding systematic uncertainty between bins of the event-shape observables shown for (upper left) transverse thrust, (upper right) isotropy, (lower left) particle multiplicity, and (lower right) invariant mass.}
\label{fig:unfold-corrsys-1d-4}
\end{figure*}

Figures~\ref{fig:unfold-corrsys-1d-3} and \ref{fig:unfold-corrsys-1d-4} provide the correlation matrices for the systematic uncertainties in the histograms given in Figs.~\ref{fig:unfold-gen-1d-1}, \ref{fig:unfold-gen-1d-2}, and \ref{fig:unfold-gen-1d-3}.
\clearpage
\section{Uncertainty decomposition}\label{sec:uncertainty_decomposition}
Table~\ref{table:uncertainty_decomposition} shows the uncertainty decomposition of unfolding, quantified by ratios of the uncertainties from individual sources to the total uncertainties, averaged over bins of transverse sphericity.
\begin{table}[htbp!]
    \centering
    \topcaption{Relative uncertainties of unfolding from different sources, averaged over bins of transverse sphericity. The uncertainty from regularization and that from the model deficiency of the migration are estimated from the square root of the quadrature sum of the differences between the unfolding result using the nominal MC simulation and those using reweighted MC samples to systematic variations. The MC statistical uncertainty is estimated from the statistical uncertainty of the histogram of weighted MC samples after the unfolding. The data statistical uncertainty is estimated from the standard deviations of the unfolding results to alternative data samples with weight following a Poisson(1) distribution. The average relative uncertainties for the distributions of other observables are similar in size.}
    \begin{scotch}{lc}
        Uncertainty source & Relative uncertainty\\
        \hline
        Track reconstruction efficiency & 1.7\%\\
        Model deficiency of the migration & 1.6\%\\
        Bias from regularization & 1.2\%\\
        MC event count & 0.26\%\\
        Data event count & 0.26\%\\
    \end{scotch}
    \label{table:uncertainty_decomposition}
\end{table}

\cleardoublepage \section{The CMS Collaboration \label{app:collab}}\begin{sloppypar}\hyphenpenalty=5000\widowpenalty=500\clubpenalty=5000
\cmsinstitute{Yerevan Physics Institute, Yerevan, Armenia}
{\tolerance=6000
V.~Chekhovsky, A.~Hayrapetyan, V.~Makarenko\cmsorcid{0000-0002-8406-8605}, A.~Tumasyan\cmsAuthorMark{1}\cmsorcid{0009-0000-0684-6742}
\par}
\cmsinstitute{Institut f\"{u}r Hochenergiephysik, Vienna, Austria}
{\tolerance=6000
W.~Adam\cmsorcid{0000-0001-9099-4341}, J.W.~Andrejkovic, L.~Benato\cmsorcid{0000-0001-5135-7489}, T.~Bergauer\cmsorcid{0000-0002-5786-0293}, S.~Chatterjee\cmsorcid{0000-0003-2660-0349}, K.~Damanakis\cmsorcid{0000-0001-5389-2872}, M.~Dragicevic\cmsorcid{0000-0003-1967-6783}, P.S.~Hussain\cmsorcid{0000-0002-4825-5278}, M.~Jeitler\cmsAuthorMark{2}\cmsorcid{0000-0002-5141-9560}, N.~Krammer\cmsorcid{0000-0002-0548-0985}, A.~Li\cmsorcid{0000-0002-4547-116X}, D.~Liko\cmsorcid{0000-0002-3380-473X}, I.~Mikulec\cmsorcid{0000-0003-0385-2746}, J.~Schieck\cmsAuthorMark{2}\cmsorcid{0000-0002-1058-8093}, R.~Sch\"{o}fbeck\cmsAuthorMark{2}\cmsorcid{0000-0002-2332-8784}, D.~Schwarz\cmsorcid{0000-0002-3821-7331}, M.~Sonawane\cmsorcid{0000-0003-0510-7010}, W.~Waltenberger\cmsorcid{0000-0002-6215-7228}, C.-E.~Wulz\cmsAuthorMark{2}\cmsorcid{0000-0001-9226-5812}
% ...
\par}
\vskip\cmsinstskip
\dag:~Deceased\\
$^{1}$Also at Yerevan State University, Yerevan, Armenia\\
$^{2}$Also at TU Wien, Vienna, Austria\\
% ...
\end{sloppypar}
\end{document}